\documentclass[12pt, draftclsnofoot, onecolumn]{IEEEtran}  
\usepackage[top=1.9cm,bottom=2.29cm,left=1.29cm, right=1.29cm]{geometry} 

\makeatletter
\def\ps@headings{%
\def\@oddhead{\mbox{}\scriptsize\rightmark \hfil \thepage}%
\def\@evenhead{\scriptsize\thepage \hfil \leftmark\mbox{}}%
\def\@oddfoot{}%
\def\@evenfoot{}}
\makeatother 
\usepackage{balance}
\usepackage{times}
\usepackage{cite}
\usepackage{float} 
\usepackage{graphicx}
\usepackage{subfigure}
\usepackage{color}
\usepackage{ifpdf}
\usepackage{epsfig}
\usepackage{latexsym}
\usepackage{amsfonts}
\usepackage{amssymb}
\usepackage{paralist}
\usepackage{comment}
\usepackage{xspace}
\usepackage{mathrsfs}
\usepackage{amssymb}
\usepackage{setspace}
\usepackage{color}
\usepackage[small,labelsep=period]{caption}

\usepackage{algorithmic}
\usepackage[ruled]{algorithm}
\usepackage{algorithmic}

\usepackage{amssymb}
\usepackage{url,epsfig,array}
\usepackage{leftidx}
\usepackage{amsmath}
\def\ie{\textit{i.e.}\xspace}
\def\etal{\textit{et al.}\xspace}

\newtheorem{theorem}{Theorem}

\newtheorem{definition}{Definition}

\def\IEEEkeywordsname{Index Terms}
\begin{document}
\title{BRAINS: Joint Bandwidth-Relay Allocation in Multi-Homing Cooperative D2D Networks\thanks{This manuscript has been accepted by IEEE Transactions on Vehicular Technology. This is not the final version. Contents may be different from the final version. For the final version, see IEEE Xplore Digital Library. Copyrights are hold by authors and corresponding copyrights holders.}}
\author{\IEEEauthorblockN{\IEEEauthorrefmark{1}Long Chen\thanks{\IEEEauthorrefmark{1}Principal corresponding author.}, \IEEEauthorrefmark{2}Jigang Wu\thanks{\IEEEauthorrefmark{2}Corresponding author.}, \textit{Member, IEEE}, \IEEEauthorrefmark{3}Hong-Ning Dai, \textit{Senior Member, IEEE} and \IEEEauthorrefmark{4}Xiaoxia Huang, \textit{Member, IEEE} \\}
\IEEEauthorblockA{\IEEEauthorrefmark{1}\IEEEauthorrefmark{2}
School of Computer Science and Technology, Guangdong University of Technology, China\\
\IEEEauthorrefmark{1}lonchen@mail.ustc.edu.cn,\IEEEauthorrefmark{2}asjgwucn@outlook.com\\}
\IEEEauthorblockA{\IEEEauthorrefmark{3}Faculty of Information Technology, Macau University of Science and Technology, Macau SAR, hndai@ieee.org}\\
\IEEEauthorblockA{\IEEEauthorrefmark{4}College of Information Engineering, Shenzhen University, xiaoxiah@gmail.com}}
\maketitle

\begin{abstract}
As an emerging technique, cooperative device to device (CD2D) communication has been considered to be a solution to capacity shortage problem. Combining multi-homing and CD2D techniques together can potentially improve network performance. We propose a novel multi-homing CD2D (MH-CD2D) network, in which multiple homing mobile devices (MMDs) act as relays for the cooperative communications of ordinary mobile devices (OMDs). We formulate such joint bandwidth-relay allocation problem as a two-stage game, in order to deal with two challenges: 1) how to design incentive mechanisms motivating MMDs to lease spare bandwidths to OMDs; 2) how to help OMDs to choose appropriate MMD relays. In the first stage, we use a non-cooperative game to model the competition between MMDs in terms of shared bandwidth and price. In the second stage, we model the behavior of OMDs selecting MMDs by an evolutionary game. We prove that there exists Nash equilibrium in the game and propose a distributed incentive scheme named IMES to solve the joint bandwidth-relay allocation problem. Extensive simulation results show that the equilibrium can be achieved and the best response price of one MMD increases with the other's best price in the Stackelberg game. The utility of MMDs increases with the number of OMDs in each OMD group at the evolutionary equilibrium. The proposed algorithms are able to reduce average service delay by more than $25\%$ in comparison to the randomized scheme which is frequently used in the most existing works. On average, IMES outperforms existing scheme by about $20.37\%$ in terms of utility of MMDs.
\end{abstract}

\section{Introduction}\label{sec:intro}
With the proliferation of various wireless devices and numerous mobile applications, there is a tremendous growth on user demands for high data rates of wireless networks. Device-to-Device (D2D) communication \cite{tehrani2014device} as an emerging communication technology is of great potential to alleviate this capacity-shortage problem by offloading traffic from base stations. Besides, cooperative communication \cite{shi2008optimal} is another new type of wireless communication technology, in which devices can help each other to relay information and consequently improve the spatial diversity. As a result, the network capacity can be enhanced. The integration of D2D communications and cooperative communications can better improve the network capacity. Recently, it is shown in \cite{chen2014exploiting} and \cite{wang2015exploiting} that cooperative D2D (CD2D) communications can improve the performance of both D2D communications and the infrastructure communications at base stations. However, most of current studies on D2D communications, cooperative communications and CD2D communications only consider (i) homogeneous network settings and (ii) associations of devices to a single network. Both considerations (i) and (ii) are neither practical nor applicable to the heterogeneous and multi-accessing features of next generation networks \cite{DhillonJSAC12}\cite{XShen:IEEEComMag15}.

Instead of associating to a single network, a device with multi-homing capability can maintain multiple simultaneous associations to several different wireless networks \cite{shi2004performance}. The multi-homing communication technology has the following advantages: (1) supporting applications with high data rate demands by aggregating bandwidth resources from multiple access networks; (2) being capable to relay information across different networks with the mobility of devices. Essentially, the multi-homing technology is mainly designed for heterogeneous networks \cite{ismail2012decentralized} since devices with multi-homing capability shall be able to travel across different access networks.

Using multi-homing technology in CD2D networks can potentially improve the network performance further. Therefore,  we propose a novel network architecture that integrates both the two technologies. We name such multi-homing CD2D networks as MH-CD2D networks. In this new network architecture, we need to solve the following research challenges:
\begin{itemize}
\item \emph{Challenge I.} From the perspective of MMDs, serving more OMDs means more energy consumption. To save energy, some selfish MMDs may decline to relay information. Essentially, we shall design appropriate incentive mechanisms to motivate MMDs to make their contributions. For example, each OMD must pay for the service offered by the corresponding MMD.
\item \emph{Challenge II.} MMDs may compete with each other for earning more profits from OMDs. However, competitions between MMDs can also cause utility losses for MMDs when charging price becomes lower and lower. Therefore, we shall coordinate bandwidth leases of MMDs.
\item \emph{Challenge III.} As the number of OMDs is usually larger than that of MMDs, every source node (OMD) always wants to choose the best relay to maximize the performance. The inefficient selection of relays may lead to the network congestion and consequently degrades the performance of CD2D communications. Therefore, an efficient bandwidth allocation and relay selection mechanism shall be designed in our MH-CD2D networks.
\end{itemize}

Recently, an interesting work was proposed in \cite{DGamageIEEEWCOM14} focusing on mode selection, but only working for two types of networks, \ie WLANs and LTE-A cellular networks. Some significant works \cite{cao2015cooperative,li2014capacity,chen2015optimal,li2014fine,zhao2015joint} focus on bandwidth allocation or relay selection in CD2D networks by assuming that relays are willing to cooperate. The incentive schemes presented in \cite{Niyato2009Dynamics} and \cite{niyato2009dynamicsCRN} utilize evolutionary game to allocate radio spectrum resources without considering relay influence. It is observed that, multiple networks are not considered in \cite{DGamageIEEEWCOM14} and incentive mechanisms are not addressed in \cite{DGamageIEEEWCOM14, cao2015cooperative,li2014capacity,chen2015optimal,li2014fine,zhao2015joint }. Moreover, spatial diversity is not included in \cite{Niyato2009Dynamics} and \cite{niyato2009dynamicsCRN}. This motivates us to design an incentive mechanism that combines multiple homing and cooperative communication techniques to solve the above mentioned challenges. This paper is distinct from our previous work \cite{lchenhpcc16} in many aspects. Firstly, our work \cite{lchenhpcc16} has been significantly extended by adding investigations for the best strategies of pricing and bandwidth allocation on both MMDs, together with comparisons between proposed algorithm and randomized scheme in terms of service delay. Moreover, the relationship between average service delay and the number of OMDs is analyzed in this paper. The main contributions of this paper are summarized as follows:

\begin{itemize}
\item We formally identify an MH-CD2D network architecture that characterizes the features of multi-homing capability and cooperative communications. The incentive mechanism for bandwidth allocation and relay selection in MH-CD2D networks has not been investigated before.
\item We model the joint bandwidth-relay allocation problem in MH-CD2D networks as a two-stage game which consists of two sub-games, (A) a non-cooperative game among MMDs; (B) an evolutionary game among OMDs. Sub-game (A) is the leader and sub-game (B) is the follower.
\item To solve the aforementioned challenges (Challenges I, II, III), we propose a distributed scheme named \textbf{i}ncentive \textbf{m}echanism for \textbf{e}fficient bandwidth relay \textbf{s}haring (IMES) in MH-CD2D networks. We also show that IMES converges fast and it is efficient for resource sharing in MH-CD2D networks.
\item To the best of our knowledge, this is the first work to investigate price competition between MMDs. Extensive simulation results show that, the average service delay can be reduced by more than $25\%$ in comparison to the randomized scheme which is frequently used in the most existing works. On average, IMES outperforms existing \textbf{c}ombinatorial \textbf{d}ouble \textbf{a}uction \textbf{r}esource \textbf{a}llocation (CDARA) scheme \cite{samimi2016combinatorial} by about $20.37\%$ in terms of utility of MMDs.
\end{itemize}

The rest of this paper is organized as follows. Section \ref{sec:new} describes our new architecture and presents basic assumptions. Section \ref{sec:related} gives a brief review of related works on both non-incentive based and incentive based resource allocation schemes. In Section \ref{sec:network}, we describe network model and define resource allocation problem. Section \ref{sec:problem} formulates the problem as a two-stage Stackelberg game. We then propose the distributed IMES scheme in Section \ref{sec:imes}. Section \ref{sec:numirical} presents the performance evaluation results and finally Section \ref{sec:conclude} concludes this paper.
\section{A New Architecture for Cooperative D2D Communication}\label{sec:new}

\begin{figure}[htb]
	\centering
	\setlength{\belowcaptionskip}{-1em}
	\includegraphics[width=3.5in]{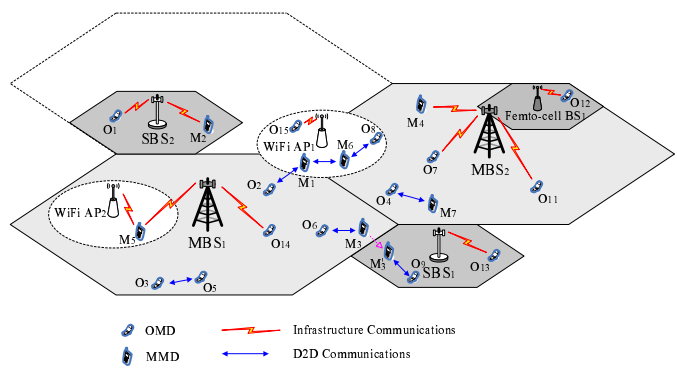}
	\caption{An example of MH-CD2D network architecture}\label{fig:system}
\end{figure}

In this architecture, some mobile devices can be associated to multiple heterogeneous networks and others can only be associated to a single network, which has following characteristics.
\begin{itemize}
\item There are two kinds of devices in MH-CD2D networks: (1) ordinary mobile devices (OMDs) and (2) multi-homing mobile devices (MMDs). OMDs are only associated to a single network or can communicate with one source node and MMDs can be associated to multiple networks. Besides, MMDs can travel across different networks while OMDs cannot\footnote{This assumption is reasonable because we consider each OMD is equipped with a half duplex antenna and a single network interface due to constrained economic budget.}.
\item MH-CD2D networks consist of multiple heterogeneous networks with different coverage area and different bandwidth settings.
\item There are two kinds of communications in MH-CD2D networks: (i) infrastructure communications and (ii) D2D communications. In infrastructure communications, both OMDs and MMDs can communicate with base stations or APs directly. In D2D communications, OMDs and MMDs can directly communicate with each other.
\end{itemize}

Fig. \ref{fig:system} shows an example of our MH-CD2D networks, which consists of macro cellular networks, small cellular networks, femto cellular networks and WLANs. In each network, both OMDs and MMDs can communicate with their associated base stations through infrastructure communications. For example, as shown in Fig. \ref{fig:system}, OMDs O$_{11}$, O$_7$ and an MMD M$_4$ can communicate with their associated macro base station, \ie, (MBS${_2}$) via infrastructure communications. Different from an OMD, an MMD can be associated with more than one network. For example, an MMD M$_5$ can be associated with MBS$_1$ and WiFi AP$_2$ simultaneously. In addition to infrastructure communications between devices and base stations, there are D2D communications in our MH-CD2D networks. Take Fig. \ref{fig:system} as an example again. OMDs O$_3$ and O$_5$ can directly communicate with each other and an OMD O$_4$ can communicate with an MMD M$_7$ in D2D mode. Besides, MMDs can travel across different networks. For example, M$_3$ can move from a macro cellular network to another small cellular network (the new position is denoted by M$^{'}_3$ and the moving direction is indicated by the dash arrow). Note that there are usually more OMDs than MMDs in our MH-CD2D networks because an MMD with multi-homing capability (e.g., equipped with multiple communication modules \cite{XShen:IEEEComMag15}) is more expensive than an OMD.

In this paper, we investigate cooperative communications in MH-CD2D networks to improve the network performance. We choose MMDs as relaying nodes instead of OMDs because: (1) MMDs can aggregate bandwidth resources from multiple access networks; (2) MMDs can help to relay information across different networks by exploiting multi-homing mobility of MMDs; (3) the case of choosing OMDs as cooperative relays can be regarded as a special case of our analysis.
\section{Related Work}\label{sec:related}
In this section, we briefly survey related studies on the joint bandwidth-relay allocation in cooperative communications, D2D communications and CD2D communications.
\subsection{Non-Incentive-Based Schemes}
\label{subsec:nonincentive}
There are many approaches proposed to improve capacity of the wireless network. Cooperative communication \cite{shi2008optimal}\cite{Nosratinia:IEEECommMag04} has received extensive attentions recently since it can boost the capacity by exploiting spatial diversity with multiple relaying nodes. Besides, D2D communications can significantly enhance the capacity by establishing a path between two wireless devices directly without going through the infrastructure of a base station or an access point. It is shown in \cite{chen2014exploiting} and \cite{wang2015exploiting} that integrating both cooperative communications and D2D communications together can improve the capacity further. We call such cooperative D2D communications as CD2D communications.

One of the key issues in CD2D communications is the \emph{joint bandwidth-relay allocation} problem. However, many of current studies address bandwidth (spectrum) allocation or relay allocation separately. For example, Cao \etal \cite{cao2015cooperative} provided a spectrum sharing framework between cellular users and D2D users in the up-link channel. An auction based relay allocation and sharing scheme was proposed by Chen \etal \cite{chen2015optimal} where cellular users helped the transmission of D2D pairs. In \cite{ayoughi2015optimized}, a relay based MIMO transmission with interference mitigation scheme was proposed for D2D communications. Multi-relay diversity was proposed by Lv \etal \cite{lv2013low} and round-robin based relay sharing was investigated in \cite{zhang2013round}.

The joint bandwidth-relay allocation in CD2D communications has attracted extensive attentions recently since it can fully utilize the resource and better improve the performance. Li and Guo \cite{li2014capacity} formulated the joint problem as a mixed-integer non-linear programming and a centralized coding-based algorithm was employed to achieve high max-min transmission rate. In \cite{li2014fine}, Li \etal proposed a fine grained resource allocation scheme with CD2D communication in cellular networks. Zhao \etal \cite{zhao2015joint} investigated a coding scheme to maximize total transmission rate in CD2D transmissions. However, most of those studies only consider an idealistic scenario, in which relay nodes are always willing to serve. It is impractical in real life as most mobile nodes are resource-constrained (e.g., power-constrained). Serving other nodes will result in the increase of self energy cost. Besides, most of the above studies fail to consider the heterogeneity of mobile nodes as all nodes are regarded as homogeneous.

\subsection{Incentive-Based Schemes}
\label{subsec:incentive}

There are some incentive-based mechanisms proposed for relay allocation in cooperative communication network\cite{yang2011truthful,wang2009distributed,al2013joint}. Yang \etal \cite{yang2011truthful} designed a McAfee based truthful double auction scheme for cooperative communications, but one relay can only serve one source node. Besides, Wang \etal \cite{wang2009distributed} presented a Stackelberg game for joint relay selection and power control, but they employed only one source-destination pair, which cannot be applied to our MH-CD2D networks where multiple MMDs act as relays. Although there were multiple source-destination pairs sharing one particular relay in \cite{al2013joint}, it is not suitable for multiple relays' scenario. Other auction schemes for spectrum sharing can refer to \cite{chen2015spectrum}. Most of those studies do not consider relay competitions.

As an efficient resource allocation methodology, Game theory \cite{fundenberg1991game} has shown its advantages in providing incentives for network users to participate in resource allocation of wireless networks. Niyato and Hossain \cite{Niyato2009Dynamics} firstly modeled the competitions among users in different areas as an evolutionary game, which was further extended to model the price competitions between primary users in a cognitive radio network in \cite{niyato2009dynamicsCRN}. However, both \cite{Niyato2009Dynamics} and \cite{niyato2009dynamicsCRN} failed to utilize cooperative spectrum diversity, which can significantly improve the capacity \cite{shi2008optimal}. Wang \etal \cite{wang2015truthful} addressed the joint relay and spectrum allocation using auction theory and Li \etal \cite{li2011coalitional} studied the coalitional game for spectrum access in cognitive radio networks while both of them did not consider relay competitions. Recently, Yun \etal \cite{yun2015energy} proposed a Stackelberg game-based scheme to tackle the network radio bandwidth allocation problem for multi-homing devices. However, they failed to address price competitions and the scenario in this paper is distinct from that in \cite{yun2015energy}. In summary, those game-theoretical solutions can not be applied to our MH-CD2D networks since they considered neither MMD competitions nor cooperative spectrum diversity. Therefore, it is the goal of this paper to address the above issues.
\section{System Model}\label{sec:network}
\begin{table}[t]
	\centering
	\small
	\caption{Basic Notations for System Model}
	\label{tb:basic}
	\begin{tabular}{lp{7cm}lp{7cm}|}
		\hline
		Notations     & Meaning                               \\ \hline
		$\mathcal{R}$ & The set of active relay MMDs          \\ \hline
		$M$           & Total number of MMDs                  \\ \hline
		$r_i$         & The i-th element of set $\mathcal{R}$ \\ \hline
		$m$           & The total number of active MMDs, $m=|\mathcal{R}|, m\le M$        \\ \hline
		$\mathcal{N}$ & The set of active OMDs, $n=|\mathcal{N}|, n\le N$                 \\ \hline
		$n_i^g$       & Number of OMDs in community $g$ that is attached to relay $r_i$ \\ \hline
		$\omega_i$    & Total bandwidth that MMD relay $r_i$ is willing to lease      \\ \hline
		$n_i$         & Total number of OMDs that choose to associate with MMD $r_i$  \\ \hline
		$S_i$         & The set of OMDs that choose MMD $r_i$                 \\ \hline
		$s_{ij}$        & The $j$th OMD that chooses relay $r_i$, $j \in [1,n_i]$                \\ \hline
		$\mathbf{P}$  & Charging price vector of all MMDs \\ \hline
		$p_{r_i}$    & Charging price of the $i$th MMD, $i\in [1,m]$ \\ \hline
		$\mathbf{\Omega}$ & The bandwidth vector that the MMDs are willing to share \\ \hline
		$d_{ij}$     & The destination node of source node $s_{ij}$ \\ \hline
		$P_{s_{ij}}$ & Transmission power of source node $s_{ij}$ \\ \hline
		$P_{r_i}$    & Transmission power of relay node $r_i$      \\ \hline
		$C_D$ & The achievable capacity between source $s_{ij}$ and destination $d_{ij}$ \\ \hline
		$C_R$ & The achievable capacity between $s_{ij}$ and $d_{ij}$ with cooperative relay $r_i$ \\ \hline
		$u_{ij}^g$   & Utility of the $j$th source $s_{ij}$ renting MMD $r_i$  \\ \hline
		$U_i$        & Utility of the $i$th MMD relay $r_i$    \\ \hline
		$\bar{u}^g$  & Average utility of a user in group $g$   \\ \hline
		$\pi_{i}^{g}$   & Fraction of sharing for relay $r_i$, $\sum_i \pi_{i}^{g}=1$, $\pi_i^g \ge 0  $ \\ \hline
 	\end{tabular}
\end{table}

We consider the overlay paradigm for the spectrum allocation between D2D communications and infrastructure communications, where the mutual interferences between the two types of communications can be ignored \cite{Goldsmith:PIEEE09}. Besides, mode selection in MH-CD2D networks is not considered since we mainly concentrate on the design of incentive mechanisms. We assume that the spectrum owned by MMDs is coordinated by network service providers, \ie base stations using OFDMA \cite{cao2015cooperative} or TDMA  \cite{zhang2014social}\cite{zhaosocial} so that interferences among MMDs and interferences within OMD communities can be mitigated. Meanwhile, by adopting network coding schemes \cite{simoni2016buffer}, we can properly handle the inter-relay interference. In the network, we assume that nodes do not frequently move or move slowly, similar to that in \cite{Fu2009Learning}. This work focuses on the type of static cases of nodes, rather than on the cases with different mobility patterns, such as random walk, hot spot mobility, route mobility models etc. \cite{lu2005study}, although the mobility of nodes is one of the most important research cases. For cost efficient location update and paging management in personal communication service networks, one may refer to the analysis performed by Li et al.\cite{li2000optimal}\cite{Li2002Analysis}. For throughput-delay-mobility tradeoff study under a more practical restricted random mobility model, one may refer to \cite{Li2010Throughput}. Basic notations are summarized in Table \ref{tb:basic}.

There are $M$ MMDs and $N$ OMDs in an MH-CD2D network. We use $\mathcal{R}=\{r_1,r_2,\cdots,r_m\}$ to denote the set of active relaying MMDs and $\mathcal{N}=\{1,2,\cdots,n\}$ to denote the set of active OMDs, where $|\mathcal{R}| \leq M$ and $|\mathcal{N}|\leq N$. Since both MMDs and OMDs are carried by mobile users who may have common interests using social applications, such as watching videos or browsing pictures, each OMD may choose to associate with multiple MMDs sharing common interests. Note that at the same time, one OMD can only be served by one MMD. That's because each OMD  can only communicate with one MMD at a time with a half-duplex communication module. Differently, each MMD is equipped with multiple antennas, which allows the MMD associating with multiple networks. Generally, an MMD may not always be busy transmitting in high data rate (e.g., video streaming). Sometimes, it performs low data rate communication (e.g., sending short messages) and only occupies a small portion of the total spectrum. In particular, an MMD can lease unused spectrum to OMDs within its community so that it can make profits while OMDs can utilize the leased spectrum via spectrum aggregation \cite{Zhang2015LTE} offered by the MMDs.

We name the OMDs associating an MMD as a \emph{community} or a \emph{group} interchangeably throughout this paper. Each community forms a cluster where the OMDs can share the spectrum and cooperative transmission opportunity of their associated MMD. It is worth mentioning that OMDs in one community can also associate with other communities, but can only be attached to one community at a time due to the half-duplexity of OMDs. In particular, we divide $n$ OMDs into a set $\mathcal{G}$ of communities (or groups), $\mathcal{G}=\{1,2.\cdots,G\}$. Apparently, we have $|\mathcal{G}|\le|\mathcal{R}|$.

In MH-CD2D networks, we assume that the total bandwidth in the system is $TB$, and let $\omega_i$ be the total bandwidth that MMD relay $r_i$ is willing to lease and let $n_i$ be the total number of OMDs that choose to associate with MMD $r_i$. We have $TB\ge \sum_i \omega_i+IB$, where $IB$ is reserved bandwidth used for infrastructure communications. Usually, $\omega_i$ has an upper bound $\bar{\omega_i}$, \ie, $0\le\omega_i\le\bar{\omega_i}$, $i\in \mathcal{R}$. Without loss of generality, we assume that OMDs in the same community are allocated with equal portion of bandwidth, which is $\omega_i/n_i$. Let $n_i^{g}$ denote the number of OMDs in community $g$ that is attached to relay $r_i$, where $g\in \mathcal{G}$. Then we have $n_i=\sum_{g} n_i^{g}$.
\section{Analysis on Joint Bandwidth-Replay Allocation in MH-CD2D Networks}\label{sec:problem}
In this section, we model the joint bandwidth-relay allocation problem in MH-CD2D networks as a two-stage Stackelberg game. In particular, for each OMD in the community, its aim is to determine which MMD to attach. Since all the OMDs are selfish, they only concentrate on their own benefits from the MMD without caring for others. As a result, multiple OMDs may choose the same relay to maximize their Quality of Service (QoS) such as the throughput capacity, which may lead to network congestion due to simultaneous connections at the same MMD. Besides, the spare spectrum resource offered by the other MMDs will be wasted. On the other hand, from the perspective of the relay, serving more OMDs means more energy consumption at MMDs and may lead to self performance degradation. To solve the above issues, we propose a \emph{market-based pricing scheme}. In this scheme, OMDs should pay for the cooperative relay service offered by MMDs. MMDs can get monetary compensation for sharing resources as a reward. Because OMDs are allocated with the same portion of bandwidth, we assume that each OMD is charged for the same price $p_{r_i}$ by the MMD $r_i$ it attaches to. Let $\mathbf{P}=\{p_{r_1},p_{r_2},\cdots,p_{r_m}\}$ denote the charging price vector by all MMDs. Let $\mathbf{\Omega}=\{\omega_1,\omega_2,\cdots,\omega_m \}$ be the bandwidth vector that the MMDs are willing to share. Then, the strategy profile for MMDs is $(\mathbf{\Omega},\mathbf{P})$. For relay set $\mathcal{R}$, the spare resources are sold to the OMDs for profits and the OMDs are buyers in the virtual market. All users in the virtual market are selfish and they only care about their own profits. The MMDs may compete with each other on the available bandwidth to offer with charging prices for OMDs. For the OMDs, each user needs to determine which MMD to employ, based on the given resource and charging profile $(\mathbf{\Omega},\mathbf{P})$. Since the OMDs are selfish, there are competitions between the users. Therefore, the competition behaviors can be formulated as a two-stage Stackelberg game, where the leaders are MMDs and the followers are OMDs.

We first describe the evolution dynamics of OMDs in subsection \ref{subsec:evolution}. Then we analyze the competition among MMDs in subsection \ref{subsec:competition}. Finally, we prove that the system will reach the Nash equilibrium in subsection \ref{subsec:simple}.
\subsection{Evolution Dynamics of OMDs}\label{subsec:evolution}
\begin{figure}[t]
	\centering
	\setlength{\belowcaptionskip}{-0.5em}
	\includegraphics[width=2.5in]{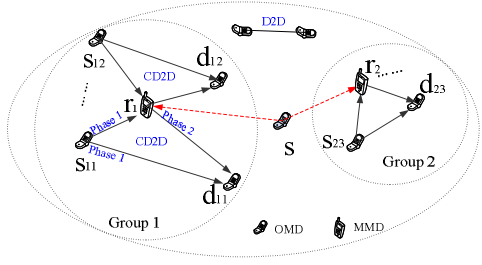}
	\caption{Demonstration of Cooperative D2D Communication}\label{fig:cc}
\end{figure}
In an MH-CD2D network, we adopt Amplify-and-Forward (AF) cooperative communication \cite{shi2008optimal}. Note that we do not consider Decode-and-Forward (DF) cooperative communication since the relay node selection algorithm designed for AF can be easily extended for DF as indicated in \cite{shi2008optimal}. AF consists of two phrases, (Phase 1) broadcasting and (Phase 2) relaying, which will be described in details as follows.

Let $S_i=\{s_{i1},s_{i2},\cdots,s_{in_i}\}$ denote the set of OMDs that choose MMD $r_i$ as cooperative relay node. Each OMD $s_{ij}\in S_i, j\in \{1,2,\cdots,n_i \}$ belongs to a group $g\in G$. Denote the destination node of source node $s_{ij}$ by $d_{ij}$. Fig. \ref{fig:cc} shows an example, in which OMDs $s_{11}$ and $s_{12}$ both choose MMD $r_1$ as cooperative relay in group $1$ while $d_{23}$ is the destination node of $s_{23}$ in group $2$. We assume that one source node can only select one MMD relay at one time due to the half-duplexity of wireless interface. For example, at one time slot, node $s$ can only be attached to $r_1$, but at next time slot, it may choose $r_2$ in group $2$.

In \emph{Phase 1}, source node $s_{ij}$ broadcasts its information $x$ to its destination node $d_{ij}$ and its information has been overheard by relay node $r_i$. The received signals $y_{s_{ij}d_{ij}}$ and $y_{s_{ij}r_{i}}$ at OMD $d_{ij}$ and MMD $r_{i}$ can be expressed as,
\begin{equation}\label{eq:rev_sd}
y_{s_{ij}d_{ij}}=\sqrt{P_{s_{ij}}}h_{s_{ij}d_{ij}}x+\xi_{s_{ij}d_{ij}},
\end{equation}
and
\begin{equation}\label{eq:rev_sr}
y_{s_{ij}r_{i}}=\sqrt{P_{s_{ij}}}h_{s_{ij}r_{i}}x+\xi_{s_{ij}r_{i}},
\end{equation}
where $P_{s_{ij}}$ is the transmission power of source node $s_{ij}$, $h_{s_{ij} d_{ij}}$ is the channel power gain between source $s_{ij}$ and destination $d_{ij}$, $h_{s_{ij} r_i}$ is channel power gain between source $s_{ij}$ and relay $r_i$. $\xi_{s_{ij}d_{ij}}$ and $\xi_{s_{ij}r_{i}}$ are additive white Gaussian noises (AWGNs) with zero mean and variance $\sigma^2$. For direct D2D transmission between $s_{ij}$ and $d_{ij}$ if relay $r_i$ is not used, the achievable capacity can be defined as
\begin{equation}\label{eq:sd}
C_D(s_{ij},d_{ij})=\frac{\omega_{i}}{\sum_{g\in G}n_i^{g}}\log_2 (1+\frac{P_{s_{ij}}|h_{s_{ij}d_{ij}}|^2}{\sigma^2}),
\end{equation}
where $SNR_{s_{ij}d_{ij}}$ is
\begin{equation}\label{eq:SNRsdij}
SNR_{s_{ij}d_{ij}}=\frac{P_{s_{ij}}|h_{s_{ij}d_{ij}}|^2}{\sigma^2}.
\end{equation}

In \emph{Phase 2}, MMD $r_i$ amplifies the received signal $y_{s_{ij}r_{i}}$ and forwards it to OMD $d_{ij}$ with power $P_{r_i}$. Then received signal at $d_{ij}$ is
\begin{equation}\label{eq:yrd}
 y_{r_i d_{ij}}=\sqrt{P_{r_i}}h_{r_{i}d_{ij}}x_{r_i d_{ij}}+\xi_{r_{i} d_{ij}},
\end{equation}
where $\xi_{r_{i}d_{ij}}$ is the received white Gaussian noise at $d_{ij}$ in the second phase with zero mean and variance $\sigma^{2}$. $x_{r_i d_{ij}}$ is the normalization factor which is defined as
\begin{equation} \label{eq:xrd}
x_{r_i d_{ij}}=\frac{y_{s_{ij} r_{i}}}{|y_{s_{ij} r_{i}}|}.
\end{equation}

Combining (\ref{eq:rev_sr}), (\ref{eq:yrd}) and (\ref{eq:xrd}), we have
\begin{equation}\label{eq:newyrd}
y_{r_i d_{ij}}=\frac{\sqrt{P_{r_i}}h_{r_i d_{ij}}(\sqrt{P_{s_{ij}}}h_{s_{ij} r_{i}}+\xi_{s_{ij} r_i})}{\sqrt{P_{s_{ij}}|h_{s_{ij} r_i}|^2+\sigma^2}}+\xi_{r_i d_{ij}}.
\end{equation}

Using (\ref{eq:newyrd}), the SNR for $s_{ij}$ relayed by $r_i$ is given as follows 
\begin{equation}\label{eq:SNRsrd}
SNR_{s_{ij} r_i d_{ij}}=\frac{P_{s_{ij}}P_{r_i}|h_{s_{ij}r_i}|^2|h_{r_i d_{ij}}|^2}{\sigma^2(\sigma^2+P_{s_{ij}}|h_{s_{ij} r_i}|^2+P_{r_i}|h_{r_i d_{ij}}|^2)}.
\end{equation}

Therefore, using (\ref{eq:SNRsdij}) and (\ref{eq:SNRsrd}) when maximal-ratio combining (MRC) is used at $d_{ij}$, the achievable capacity through relay $r_i$ can be expressed as
\begin{equation}\label{eq:cc_srd}
C_R(s_{ij},r_i,d_{ij})=\frac{\omega_{i}}{2\sum_{g\in G}n_i^{g}}\log_2 (1+SNR_{s_{ij}d_{ij}}+SNR_{s_{ij}r_i d_{ij}}).
\end{equation}

It should be noted that, the channel related parameters in the above equations are maintained at the base station and broadcast to mobile nodes via periodically control packets. We then define \emph{Relay Selection Condition} to determine whether a relay should be chosen.

\begin{definition}\label{definition:relayselection}
\emph{Relay Selection Condition.} For each source OMD $s_{ij}$, it may choose relay $r_i$ as the cooperative node if the achievable capacity gain $C_R(s_{ij},r_i,d_{ij})/C_D(s_{ij},d_{ij})$ is greater than $1$. Otherwise, it may choose direct transmission without the relay or imitate other OMD's choice within the same group to maximize the utility, which is defined in (\ref{eq:U_ijg}).
\end{definition}

Since OMDs in the same group can learn from each other's choice via the broadcasting of MMD relay and then dynamically adjust their strategies, the behaviors of players (OMDs) form the evolution in the group. According to \cite{tembine2007delayed}, for each member in group $g$, the sharing vector for relay $r_i$ can be defined as the fraction of the spectrum offered by $r_i$. Let $\mathbf{\Pi}^{g}$ be the state vector of the user in group $g$, thus
\begin{equation}\label{eq:pig}
\mathbf{\Pi}^{g}=\{\pi_{1}^g, \pi_{2}^g,\cdots,\pi_{m}^g \},
\end{equation}
where each element $\pi_{i}^g$ in $\Pi^{g}$ represents the fraction of sharing for relay $r_i$. Note that $\sum_{i} \pi_{i}^g=1$ and $\pi_{i}^g\ge 0$. For example, let $G=\{1,2\}$, $m=2$, $\pi_{1}^{1}=30\%$, $\pi_{2}^{1}=70\%$ means in a system where there are two groups and two MMDs, $30\%$ members in group $1$ choose relay $r_1$ and $70\%$ members in group $1$ choose relay $r_2$. Combining (\ref{eq:sd}) and (\ref{eq:cc_srd}), we can define the utility $u_{ij}^{g}$ of the $j$th source user $s_{ij}$ renting relay $r_i$ as follows,
\begin{equation}\label{eq:U_ijg}
u_{ij}^{g}=[\alpha t_{ij}\max \{C_R,C_D\}-{p_{r_i}}]x_{ij},
\end{equation}
where $\alpha$ is the unit cost per bit, $t_{ij}$ is the time fraction that the $j$th source user $s_{ij}$ can get access to relay $r_i$ and is usually set as $1$. It should be noted that $x_{ij}$ equals to $1$ if $C_R>C_D$ and $x_{ij} $ equals to $0$, otherwise. According to \cite{tembine2007delayed}, the utility of OMD at time $t$ is a function of the group state at time $(t-\tau)$ and $\tau$ is a time delay variable. Here, we assume the OMDs within each group have similar distances to their corresponding receivers. In this way, the sub index `$j$' in (\ref{eq:U_ijg}) can be omitted. Similar to \cite{Niyato2009Dynamics}, we have the replicator dynamics of the population (\ie, the OMDs in the group) as follows,
\begin{equation}\label{eq:dotpiig}
\dot{\pi_i}^g(t)= \delta \pi_i^{g} (t) (u_i^{g}(t-\tau)-\bar{u}^{g}(t-\tau)),
\end{equation}
where $\delta$ represents the updating speed based on the observation of OMDs, $u_i^{g}$ stands for the utility of an OMD in group $g$ that attaches to MMD $r_i$, and $\bar{u}^{g}$ is the average utility of a user in group $g$.

It is shown in (\ref{eq:dotpiig}) that the number of OMDs choosing MMD $r_i$ will increase if the payoff is higher than the group's average payoff. The evolution will end when the sharing strategy remains constant, which is the \emph{evolutionary equilibrium}.

\vspace{-1em}
\subsection{Competition Between MMDs}\label{subsec:competition}
When OMDs reach an equilibrium, MMDs will compete with each other to adjust their strategies which is coordinated by the base station to maximize total utility. The behavior of MMDs can be modeled as a \emph{non-cooperative} game. The utility of one MMD is the total payment of all OMDs that attach to it minus its cost for serving as a relay. Let $U_i$ be the utility of the $i$th MMD relay $r_i$, then we have
\begin{equation}\label{eq:U_mmu}
U_i=p_{r_i} \sum_{g\in \mathcal{G}} n_i^{g}(\mathbf{\Omega},\mathbf{P}) -c_i \omega_i,
\end{equation}
where $c_i$ is the relay cost per unit bandwidth.

\begin{definition}\label{definition:1}
Define the profile strategies for all MMDs except relay $r_i$ as $(\mathbf{\Omega_{-i}},\mathbf{P}_{-i})$.
\end{definition}

\begin{definition}\label{definition:2}
The strategy profile $\left(\mathbf{\Omega}^{*},\mathbf{P}^{*} \right)$ is the Nash equilibrium if $\forall i\in \mathcal{R}$, $U_i(\omega_i^{*},\mathbf{\Omega_{-i}^{*}};p_i^{*}, \mathbf{P}_{-i}^{*}) \ge U_i(\omega_i,\mathbf{\Omega_{-i}^{*}};p_i, \mathbf{P}_{-i}^{*})$ for $\forall \omega_i \ge 0, p_i \ge 0$.
\end{definition}

The Nash equilibrium can be found in solving the fixed point of all the best responses of players.
\begin{definition}
The optimal strategy for each MMD $i$ is defined as
\begin{equation}\label{eq:OPTi}
OPT_i(\mathbf{\Omega_{-i}},\mathcal{P}_{-i})= \arg \max_{\omega_i,p_i}U_i(\omega_i,\mathbf{\Omega_{-i}};p_i,\mathbf{P_{-i}}).
\end{equation}
\end{definition}
Next, we use a simple example to illustrate how the system evolves towards the Nash equilibrium.

\subsection{Simple Example}
\label{subsec:simple}

We consider a simple example similar to \cite{Niyato2009Dynamics}, in which there are two MMDs $\mathcal{R}=\{r_1,r_2\}$. Under this setting, we will utilize backward induction to derive the Nash equilibrium.
\subsubsection{Evolution Dynamics of OMDs}
Assume that each OMD tells its utility truthfully, as illustrated in the previous section, when $\dot{\pi}_{i}^{g}(t)=0$, the evolutionary game will converge to the equilibrium. The evolution will end because no member in the group can enhance its utility by copying other's strategy. Thus, we can solve the following equation to determine the equilibrium,
\begin{equation}\label{eq:suit}
u_{i}^{g}= u_{{j \ne i}}^{g}, \forall g\in \mathcal{G},
\end{equation}
where the left hand side of (\ref{eq:suit}) is the utility of an OMD member in group $g$ that chooses MMD $r_i$ as relay node, while the right hand side of (\ref{eq:suit}) is the utility of OMD who chooses the other relays except relay $r_i$. Combining (\ref{eq:sd})(\ref{eq:cc_srd})(\ref{eq:U_ijg})(\ref{eq:suit}), then (\ref{eq:suit}) can be re-written as (\ref{eq:loglog}),
	\begin{table*}[ht]
		\small
		\begin{equation}\label{eq:loglog}
		\begin{split}
		& \frac{\omega_1}{n_1} \log_2 \max(1+\frac{P|h_{s_1 d_1}|^2}{\sigma^2}+\frac{P^2|h_{s_1 r_1}|^2|h_{r_1 d_1}|^2}{P|h_{s_1 r_1}|^2 \sigma^2+P|h_{r_1 d_1}|^2 \sigma^2+\sigma^4}, 1+\frac{P|h_{s_1 d_1}|^2}{\sigma^2})-p_{r_1} \\ =
		& \frac{\omega_2}{n_2} \log_2 \max(1+\frac{P|h_{s_2 d_2}|^2}{\sigma^2}+\frac{P^2|h_{s_2 r_2}|^2|h_{r_2 d_2}|^2}{P|h_{s_2 r_2}|^2 \sigma^2+P|h_{r_2 d_2}|^2 \sigma^2+\sigma^4}, 1+\frac{P|h_{s_2 d_2}|^2}{\sigma^2})-p_{r_2} \\
		\end{split}
		\end{equation}
		\begin{subequations}
			\begin{align}
			  A &=\max(1+\frac{P|h_{s_1 d_1}|^2}{\sigma^2}+\frac{P^2|h_{s_1 r_1}|^2|h_{r_1 d_1}|^2}{P|h_{s_1 r_1}|^2 \sigma^2+P|h_{r_1 d_1}|^2 \sigma^2+\sigma^4}, 1+\frac{P|h_{s_1 d_1}|^2}{\sigma^2})  \label{eq:A} \\
		B &=\max(1+\frac{P|h_{s_2 d_2}|^2}{\sigma^2}+\frac{P^2|h_{s_2 r_2}|^2|h_{r_2 d_2}|^2}{P|h_{s_2 r_2}|^2 \sigma^2+P|h_{r_2 d_2}|^2 \sigma^2+\sigma^4}, 1+\frac{P|h_{s_2 d_2}|^2}{\sigma^2})  \label{eq:B}
			\end{align}
		\end{subequations}
		\hrule
	\end{table*}
where we assume that OMDs and MMDs are assigned with same transmission power $P$ and same noise at the receiver for the benefit of simplicity. That's because the different power consumptions can be modeled by the variant charging prices and costs. Regarding to power control game in the fading channel, please refer to \cite{shi2009game}. In group one, the channel gain between source $s_i$ and relay $r_i$ is denoted by $h_{s_i,r_i}$, the channel gain between source $s_i$ and destination $d_i$ is denoted by $h_{s_i,d_i}$, the channel gain between relay $r_i$ and destination $d_i$ is denoted by $h_{r_i,d_i}$ where $i\in\{1,2\}$. Let $n^g$ be the number of OMDs in $g\in \mathcal{G}$, then $n^{g}=\sum_{i\in \mathcal{R}}n_i^{g}$ and $n_i^{g}=\pi_{i}^{g} n^{g}$. Therefore, (\ref{eq:loglog}) can be rewritten as,
\begin{equation}\label{eq:eq0}
DX^2-(\omega_1 A+ \omega_2 B + nD) X+ \omega_1 A n = 0,
\end{equation}
where
\begin{equation}\label{eq:X}
X=n_1
\end{equation}
and $A$ is defined in (\ref{eq:A}), $B$ is described in (\ref{eq:B}). $D$ and $n$ are defined as follows,
\begin{equation}\label{eq:D}
D={p_{r_1}-p_{r_2}},
\end{equation}
\begin{equation} \label{eq:n}
n=\sum_{g} n^g.
\end{equation}
Note that (\ref{eq:X}) is the positive solution of equation (\ref{eq:eq0}). As we can see from (\ref{eq:loglog})
$\sim$ (\ref{eq:n}), the term $\omega_i/(\sum_{g\in \mathcal{G}} n_i^g)$ makes equations not easy to be handled. Note that the $\log(\cdot)$ function is concave, $\zeta log(\cdot)$, $\zeta>0$ is also concave. To simplify the analysis, we modify the utility function of OMDs by re-defining $C_R$ and $C_D$ (see (\ref{eq:sd}) and (\ref{eq:cc_srd})) as follows,
\begin{equation}\label{eq:new_CR}
C_R(s_{ij},r_{i},d_{ij})=\log_2(\frac{k_{\omega} \omega_i}{\sum_{g\in \mathcal{G}} n_i^g} {\tau}_{ij}),
\end{equation}
and
\begin{equation}\label{eq:new_DT}
C_D(s_{ij},d_{ij})=\log_2(\frac{k_{\omega} \omega_i}{\sum_{g\in \mathcal{G}} n_i^g}b_{ij}),
\end{equation}
where $k_{\omega}$ is a system parameter which originates from the coefficient of (\ref{eq:sd}) or (\ref{eq:cc_srd}). Note that through this change, we do not change the concave feature of the original equations (\ref{eq:sd}) and (\ref{eq:cc_srd}), which is also adopted by \cite{nan2014stackelberg}, thus we have
\begin{equation}
\tau_{ij}=1+SNR_{s_{ij}d_{ij}}+SNR_{s_{ij}r_i d_{ij}}
\end{equation}
and
\begin{equation}
b_{ij}=1+SNR_{s_{ij}d_{ij}}.
\end{equation}

As the same as previous equations (\ref{eq:dotpiig})(\ref{eq:loglog}), the subindex `$j$' can be omitted. Therefore, (\ref{eq:suit}) can be re-written as,
\begin{equation}\label{eq:equilibrium}
 \sum_{g}\pi_{1}^g n^g = \frac{n}{1+2^{p_{r_1}-p_{r_2}} \frac{\omega_2 Y_2}{\omega_1 Y_1}},
\end{equation}
where $Y_1$=$\max\{\tau_1,b_1\}$, $Y_2$=$\max\{\tau_2,b_2\}$. According to \cite{niyato2009dynamicsCRN}, the equilibrium can be determined by solving the Jacobian matrix,
\begin{equation}
\mathbf{J}=\begin{bmatrix}
   \frac{\partial \delta \pi_{1}^{(1)} (u_1^{(1)}-\bar{u}^{(1)})}{\partial \pi_1^{(1)}} & \frac{\partial \delta \pi_{1}^{(1)} (u_1^{(1)}-\bar{u}^{(1)})}{\partial \pi_1^{(2)}} \\
   \frac{\partial \delta \pi_{1}^{(2)} (u_1^{(2)}-\bar{u}^{(2)})}{\partial \pi_1^{(1)}} & \frac{\partial \delta \pi_{1}^{(2)} (u_1^{(2)}-\bar{u}^{(2)})}{\partial \pi_1^{(2)}}
\end{bmatrix}
=
\begin{bmatrix}
J_{1,1} & J_{1,2} \\
J_{2,1} & J_{2,2}
\end{bmatrix},
\end{equation}
where $\delta$ is as same as that defined in (\ref{eq:dotpiig}). The two eigenvalues of $\mathbf{J}$ can be
\begin{equation}
\lambda(\mathbf{J})=\frac{(J_{1,2}+J_{2,2})
	\pm \sqrt{4J_{1,2}J_{2,1}+(J_{1,1}-J_{2,2})^2}}{2}.
\end{equation}

\subsubsection{Pricing competition between two MMDs}
When there are multiple MMDs in the system, they may compete with each other in terms of price and bandwidth to sell to OMDs. In this paper, we consider the fixed bandwidth sharing between MMD $r_1$ and MMD $r_2$. Therefore, MMDs form the non-cooperative sub-game. The strategy of each MMD is the price charging for OMDs. The payoff of each MMD is the total payment by OMDs minus the consumption for serving OMDs. In this example, the non-cooperative game between MMDs can be defined as a tuple $\langle \mathcal{R}, {(P_i)_{i\in \mathcal{R}}},{(U_i)_{i \in \mathcal{R}}} \rangle$. Because the bandwidth sharing is fixed, the utility function of the $i$th player, defined in ($\ref{eq:U_mmu}$) can be rewritten as,
\begin{equation}\label{eq:newU_i}
U_i=p_{r_i} \times \sum_{g\in \mathcal{G}} \pi_i^{g} n^g - c_i \omega_i,
\end{equation}
where the term $\sum_{g\in\mathcal{G}}\pi_{i}^g n^g$ can be determined by (\ref{eq:equilibrium}), which is the equilibrium of the evolutionary game for the OMDs. Substituting (\ref{eq:equilibrium}) into (\ref{eq:newU_i}), we have
\begin{equation}\label{eq:Ui_w_OMD}
U_i=p_{r_i} \times {\frac{n}{1+2^{p_{r_i}-p_{r_j}} \frac{\omega_j Y_j}{\omega_i Y_i}}}- c_i \omega_i,
\end{equation}
where (\ref{eq:Ui_w_OMD}) shows the interaction between the MMD and OMD. The first derivative of $U_i$ with respect to $p_{r_i}$ is computed as
\begin{equation}\label{eq:particalUiPri}
\frac{\partial{U_i}}{\partial{p_{r_i}}}= \frac{n\omega_i Y_i [\omega_i Y_i+(2^{p_{r_i}-{p_{r_j}}}) \omega_j Y_j - p_{r_i} \omega_j Y_j {2^{p_{r_i}-p_{r_j}} \ln 2}]}{[\omega_i Y_i + (2^{p_{r_i}-p_{r_j}}) \omega_j Y_j ]^2}.
\end{equation}

Therefore, the optimal price for the MMD can be calculated by letting $\frac{\partial U_i}{\partial p_{r_i}}=0$. Then from (\ref{eq:particalUiPri}), we have
\begin{equation}\label{eq:p_r_i_optimal}
p_{r_i}^{*}=\frac{1}{\ln 2}+\frac{\omega_i Y_i}{\omega_j Y_j} 2^{p_{r_j}-p_{r_i}}.
\end{equation}

In order to obtain the expression of the optimal price $p_{r_i}$, without loss of generality, we use the function $\exp(\cdot)$ to replace the function $2^{(\cdot)}$, then (\ref{eq:p_r_i_optimal}) can be rewritten as
\begin{equation}\label{eq:final_pri_opt}
p_{r_i}^{*}=\mathcal{B}_i(p_{r_j})=1+\mathcal{W}\{\frac{\omega_i Y_i}{\omega_j Y_j} \exp^{(p_{r_j}-1)}\},
\end{equation}
where $\mathcal{W}(\cdot)$ is a Lambert-W function \cite{Wackerly:2007}.

\subsubsection{Nash equilibrium for non-cooperative game between two MMDs}
According to (\ref{eq:final_pri_opt}), each MMD's optimal price increases with the growing of other MMD's price. Therefore, this game is proved to be a supermodular game \cite{niyato2009dynamicsCRN} and there is Nash equilibrium in this game.

\begin{theorem}\label{th:exists}
There is a pure Nash equilibrium in the non-cooperative game between MMDs.
\begin{proof}
 A super-modular game must satisfy the following properties: Property 1) A strategy is a subset of real set. Property 2) A utility has increasing difference in all sets of strategies. In the non-cooperative game, the optimal strategy of player $i$, its price $p_{r_i}\in \left[0,+\infty\right]$. According to \cite{niyato2009dynamicsCRN}, property $1$ can be easily verified. For property $2$, it can be verified through $\frac{\partial^2 U_i}{\partial p_{r_i} \partial p_{r_j}} \ge 0$. We can calculate $\frac{\partial^2 U_i}{\partial p_{r_i} \partial p_{r_j}}$ as
 \begin{equation}\label{eq:partial2Uiij}
\frac{\partial^2 U_i}{\partial p_{r_i} \partial p_{r_j}}=\frac{\partial \mathcal{B}_i}{\partial p_{r_j}}=\frac{\mathcal{W}\left(z\right)}{z(1+\mathcal{W}(z))},z=e^{(p_{r_j}-1)}.
 \end{equation}
Since $z>0$, and also, we have
\begin{equation}
z=\mathcal{W}(z)e^{\mathcal{W}(z)}.
\end{equation}
hence, we have $\mathcal{W}>0$ because $z>0$. Further, we have $\frac{\partial^2 U_i}{\partial p_{r_i} \partial p_{r_j}}\ge 0$. Therefore, the non-cooperative game is a supermodular game with a Nash equilibrium.
\end{proof}
\end{theorem}

It is worthy noting that, the channel conditions are known a priory at base stations and are broadcasted to each mobile devices in the network via control packets. We then show that the non-cooperative game has a unique Nash equilibrium, shown as follows.
\begin{theorem}\label{th:uniqueNE}
The above non-cooperative game has a unique Nash equilibrium.
\begin{proof}
According to Theorem \ref{th:exists}, there is a Nash equilibrium in the non-cooperative game between MMDs. In this theorem, we prove that the Nash equilibrium is unique by showing the self-mapping function of the MMD's best response function is a contraction. The self-mapping function is expressed as follows,
\begin{equation}
p_{r_i}=\mathcal{F}_i(p_{r_i})=\mathcal{B}_i(\mathcal{B}_j(p_{r_i}))=1+\mathcal{W}\left(\frac{e^{p_{r_i}-1}}{\mathcal{W}\left(\frac{\omega_j Y_j}{\omega_i Y_i} e^{p_{r_i}-1}\right)}\right)
\end{equation}
for $i\in \{1,2\}$ and $j\ne i$. To prove the uniqueness of Nash equilibrium, it is equivalent to demonstrate the self-mapping function is a contraction. We observe that the Jacobian matrix is
\begin{equation}
\mathbf{J}=\begin{bmatrix}
\frac{\partial^2 \mathcal{F}_i}{\partial p_{r_i}^2 } & \frac{\partial^2 \mathcal{F}_i}{\partial p_{r_i} \partial p_{r_j} } \\
\frac{\partial^2 \mathcal{F}_j}{\partial p_{r_i} \partial p_{r_j} } & \frac{\partial^2 \mathcal{F}_j}{\partial p_{r_j}^2 }
\end{bmatrix}=\begin{bmatrix}
0 & \frac{\partial \mathcal{B}_i}{\partial p_{r_j}} \\
\frac{\partial \mathcal{B}_j}{\partial p_{r_i}} & 0
\end{bmatrix}.
\end{equation}
Let $\lambda$ be the largest absolute eigenvalue of the Jacobian matrix $\mathbf{J}$. Then we have
\begin{equation}
\lambda = \sqrt{ \frac{\partial \mathcal{B}_i}{\partial p_{r_j}}\times \frac{\partial \mathcal{B}_j}{\partial p_{r_i}}}.
\end{equation}
The self-mapping function is a contraction if and only if $\lambda<1$. According to (\ref{eq:partial2Uiij}), we have $\frac{\partial \mathcal{B}_i}{\partial p_{r_j}}<1$ and thus $\lambda <1$ is established, which means the self-mapping function is a contraction. Therefore, the uniqueness of Nash equilibrium for MMDs is proved.
\end{proof}
\end{theorem} 
\section{IMES: A DistriBbuted Implementation}\label{sec:imes}
In this section, a distributed Incentive Mechanism for Efficient bandwidth relay Sharing in CD2D networks (IMES) is proposed. The mechanism includes evolution among OMDs and price competition within MMDs. For MMDs, they are leaders and make decisions on the price and bandwidth for OMDs through competition. The objective is to maximize their own utilities. For OMDs, they are followers and only respond to the ask price and bandwidth changes made by MMDs to maximize their own utilities.
\subsection{Evolution Mechanism Among OMDs}
In this subsection, an iterative mechanism is proposed for OMDs so that they converge to the equilibrium. Each OMD in the group tries to maximize its own utility. A distributed algorithm is designed to execute at each OMD. Within each group, OMDs can communicate with each other and share the information of the other's choice. The average utility sent by MMDs can be shared with members in each group. Each member in the group can change the decision with a probability if observing the other member within the group obtains a higher utility. The evolution will end when all OMDs in the same group achieve equal or almost equal utility. The details of the procedure are illustrated as follows. Firstly, each OMD in the group is attached to a MMD relay node randomly. Then, each MMD computes its utility by using the allocated bandwidth and the price charged by the corresponding MMD from (\ref{eq:U_ijg}). After exchanging information with users in the same group, each OMD will get the information of utility and choice to compute the group's average utility according to the following equation \cite{nan2014stackelberg},
\begin{equation}\label{eq:imes_averageuti}
\bar{u}^g(t)= \frac{\sum_i u_{i}^{g}(t-1)n_i^{g}(t-1)}{n^g (t-1)}.
\end{equation}

The OMD will change its link to the MMD if it finds another MMD can provide a higher utility when the probability $\psi (t)$ is higher than the random number within $[0,1]$, where $\psi(t)$ is defined as \cite{Niyato2009Dynamics}
\begin{equation}
\psi(t)= \frac{\bar{u}^{g}(t)- u_{i}^{g}(t)}{\bar{u}^g(t)}.
\end{equation}

The OMDs will repeat those procedures until the evolution is stable. The details of the scheme executed on each OMD are shown in Algorithm \ref{alg:OMD}.
\begin{algorithm}
	\begin{algorithmic}[1]
		\caption{Function DAO( ): Distributed Algorithm on each OMD }\label{alg:OMD}
	\STATE{For all OMDs, the $i$th MMD $r_i$ is randomly chosen.}
	\STATE{$flag=true$;}
	\WHILE{$flag$}
	\STATE{A OMD computes the utility $u_{ij}^{g}$ from the obtained price and bandwidth using (\ref{eq:U_ijg}). The utility is then shared within the group $g$.}
	\STATE{Receive the choices and utility of the other users within the same group and computes the average utility $\bar{u}^g(t)$ in group $g$ using (\ref{eq:imes_averageuti}).}
	\IF{($u_i^{g}(t)<\bar{u}^g (t)$)}
	\STATE{Leave the current link to relay if $rand < \psi(t)$.}
	\ELSE
	  \STATE{Remain the current link to current relay MMD node}
	\ENDIF
	\IF{All OMDs within the same group have the same utility}
	\STATE{$flag=false$;}
	\ENDIF
	\ENDWHILE
\end{algorithmic}	
\end{algorithm}
\begin{algorithm}
	\begin{algorithmic}[1]
		\caption{Function DAM( ): Distributed Algorithm on each MMD}\label{alg:MMD}
		\STATE{For each MMD:}
		\STATE{Initialize the bandwidth that MMD $r_i$ is willing to share $\omega_{i,0}$, charging price $p_{r_{i,0}}$.}
		\WHILE{($\omega_i$ and $p_{r_i}$ are not stable)}
		\STATE{Wait for an interval $T$ for the evolution among OMDs by executing Function DAO( )}
		\STATE{$\omega_{i} {(t+1)}=\min\{\max\{\omega_{i,0}(t+1),0\},50\}$ // Update the bandwidth}
		\STATE{$p_{r_i}(t+1)=\max\{p_{r_{i,0}}(t+1),0 \}$ // Update the price}
		\STATE{t=t+1}
		\ENDWHILE
	\end{algorithmic}	
\end{algorithm}
\subsection{Competition implementation for MMDs}
In real scenarios, the multi-homing base stations are heterogeneous, a central controller to inform MMDs may not be always available. Therefore, each MMD may learn and make its own decision to adapt to changes. Under this setting, an iterative based updating scheme is proposed. Similar to \cite{Niyato2009Dynamics}, the MMD can update strategies as follows,
\begin{equation}\label{eq:updating_omigai}
\small
\centering
\omega_{i,0}(t+1)=\omega_i(t)+\mu_{i,\omega} \frac{U_i (\mathbf{\omega}(t),\mathbf{p}(t))-U_i(\mathbf{\omega}(t-\Delta t),\mathbf{p}(t-\Delta t))}{\Delta t},
\end{equation}
and
\begin{equation}\label{eq:updating_pri}
\small
\centering
p_{r_{i,0}}(t+1)=p_{r_i}(t)+\mu_{i,p} \frac{U_i (\mathbf{\omega}(t),\mathbf{p}(t))-U_i(\mathbf{\omega}(t-\Delta t),\mathbf{p}(t-\Delta t))}{\Delta t},
\end{equation}
where $\Delta_t$ is a period of time and $\mu_{i,\omega}$, $\mu_{i,p}$ are updating speed related parameters. When the offered bandwidth and charging price are updated, they will be broadcast to OMDs within the group. After receiving those data, each OMD will adjust their own strategies. As same as \cite{nan2014stackelberg}, due to the unknown evolution time, we also denote a waiting time $T$ for the strategy updating interval. When the results in (\ref{eq:updating_omigai}) and (\ref{eq:updating_pri}) are stable, then the Nash equilibrium is achieved by approximation. Details of the algorithm on the MMD is shown in Algorithm \ref{alg:MMD}.
\section{Performance Evaluation}\label{sec:numirical}
\begin{table}[t]
	\centering
	\caption{Default simulation parameters}
	\label{tb:simupara}
	\begin{tabular}{cc}
		\hline
		Parameters for Simulation                         & Values     \\ \hline
		$\mathcal{G}$, the set of social groups with OMDs & $\{a, b\}$ \\
		$\mathcal{R}$, the set of MMD relays              & $\{r_1,r_2\}$  \\
		$\tau$, time delay parameter                      & $1$        \\
		$T$, waiting time                                & $100ms$      \\
		$\delta$, updating speed parameter                & $1$        \\
		$n^{a}$, the number of OMDs in group $a$          & $10$       \\
		$r$, the communication range of devices           & $50m$        \\
		$n^{b}$, the number of OMDs in group $b$          & $30$       \\
		Default network area size                         & $100m\times 100m$ \\
		$\alpha$, system parameter                        & $1$      \\   \hline
	\end{tabular}
\end{table}

According to \cite{an2009evolutionary}, due to the price simultaneously charged by MMDs, there is no optimal strategy for each OMD. Therefore, the equilibrium performance is studied. In this section, we first evaluate the performance of the proposed game theoretic scheme in Section \ref{subsec:best}. We then analyze the service delay in Section \ref{subsec:service_delay}.
\subsection{Experimental Setup}
The experimental evaluation has been conducted on Matlab simulator \cite{matlab2017}. To evaluate the performance of proposed group behavior of users in neighboring groups as most studies do \cite{Xiong2017Group} \cite{Han2016Exploiting}, by default we assume that there are two groups of OMDs randomly distributed in a $100m\times100m$ area \cite{Chen2017optimal} and competing for two MMDs. As did in \cite{Li2013Internode}, we consider small groups by setting the default number of OMDs in group $a$ as $10$ while the default number of OMDs in group $b$ is $30$ . During the simulation, nodes may stay static or do not frequently move \cite{Fu2009Learning} in the network. For example, in a wireless local area network covering a metropolitan area \cite{Ma2017dota}, users may only move around in a restricted area close to their homes, including the office\cite{raghunathan2009wardrop}, gymnasiums, and so on. As another example, defending soldiers are allowed to move in their defend areas on the battlefield \cite{Li2010Throughput}. The communication range is set as $50m$ for mobile devices\cite{Wang2014Delay} and the waiting time interval is set as $100ms$.

We then set system parameters for group $a$. In particular, the channel gain between source and relay is $h_{s_1 r_1}=0.3$, the channel gain between source and destination is $h_{s_1 d_1}=0.25$ and the channel gain between relay and destination is $h_{r_1 d_1}=0.4$\cite{Chen2015Primary}. Similarly, we set system parameters for group $b$ as $h_{s_2 r_2}=0.25$, $h_{r_2 d_2}=0.35$ and $h_{s_2 d_2}=0.21$. Here we adopt the same channel gain of elements in each group to simplify the analysis just the same as in \cite{wang2006grouped} and \cite{ho2012energy}, which does not affect the performance comparison for the proposed scheme. We choose the common transmission power as $P=2W$\cite{Luo2014Green}\cite{Miao2011Distributed} for group $a$ and group $b$. Without loss of generality, we set $t_{ij}=1s$\cite{Li2017Delay}. The results are obtained by averaging over 1000 simulation runs. Table \ref{tb:simupara} lists the parameters used in simulations. The behaviors of OMDs in one group are compared with those of OMDs in the other group on choosing MMD relay nodes when Algorithms \ref{alg:OMD} and \ref{alg:MMD} are executed. The best strategy of one MMD with respect to the behavior of other MMD is also studied. To evaluate service delay, we design a randomized method to compare with the proposed IMES. When there are more than two groups, \ie $|\mathcal{R}|\ge 3$, there are two cases since in the simulation we set $|\mathcal{G}|=|\mathcal{R}|$.

Case 1: The number of MMDs is odd. In this case, the algorithms will be executed between adjacent groups and neighboring MMDs. Finally three groups of OMDs and three MMDs are left. Then we can divide one group of OMD users in the last three groups into two separate subgroups uniformly, and then duplicate the role of one MMD for the separated two groups by choosing one MMD from the last three MMDs.

Case 2: The number of MMDs is even. As same as Case I, the algorithms will be executed between neighboring OMD groups with the corresponding MMDs.
\subsection{Best Strategy}\label{subsec:best}
\begin{figure}[h]
	\centering
	\includegraphics[width=2.5in]{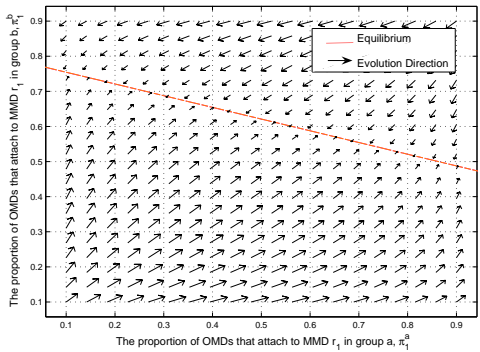}
	\caption{Phase plane of replication dynamics when $\omega_1=20$, $\omega_2=40$, $n^a=10$ and $n^b=30$}\label{fig:phaseplane}
\end{figure}

Firstly, we analyze the phase plane of replicator dynamics, as shown in Fig. \ref{fig:phaseplane}. Although there are multi-rounds for the replicator dynamics, we only consider a particular round given the offered resources and charging prices. We assume that all OMDs can perform cooperative communications with MMD $r_1$. In this simulation, $\omega_1=20$, $\omega_2=40$, $n^{a}=10$ and $n^{b}=30$. For the charging price, we set $p_{r_1}=1$ and $p_{r_2}=2$. The proportions of OMDs in group $a$ and $b$ choosing the MMD relay $r_1$ are plotted in Fig. \ref{fig:phaseplane}. The phase plane shows the evolution direction of the replicator dynamics towards the Nash equilibrium. It is shown in Fig. \ref{fig:phaseplane} that whatever the starting point is, the population of OMDs converges to different evolutionary equilibria. The solid line in Fig.\ref{fig:phaseplane} shows the set of Nash equilibria. For example, from $(\pi_1^{a},\pi_1^{b})=(0.56,0.21)$, the equilibrium will be achieved at $(\pi_1^{a},\pi_1^{b})=(0.84,0.51)$. It is worth noting that in the equilibrium, the utility of OMDs in $a$ is almost equal to that in $b$.

\begin{figure}[t]
	\centering
	\includegraphics[width=2.5in]{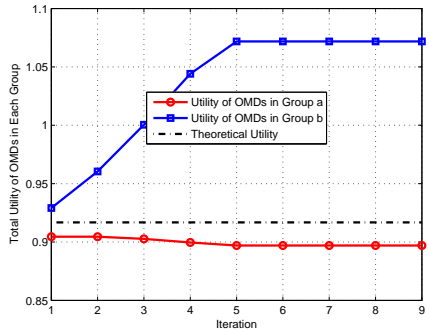}
	\caption{The utility of OMDs in two groups converge to equilibrium}\label{fig:Gutility2Nash}
\end{figure}

\begin{figure*}[htb]
	\centering
	\setlength{\abovecaptionskip}{-0.2cm}
	\setlength{\belowcaptionskip}{-1em}
	\subfigure[]{
		\label{fig:n1vsnbvp2}
		\includegraphics[width=0.2\textwidth]{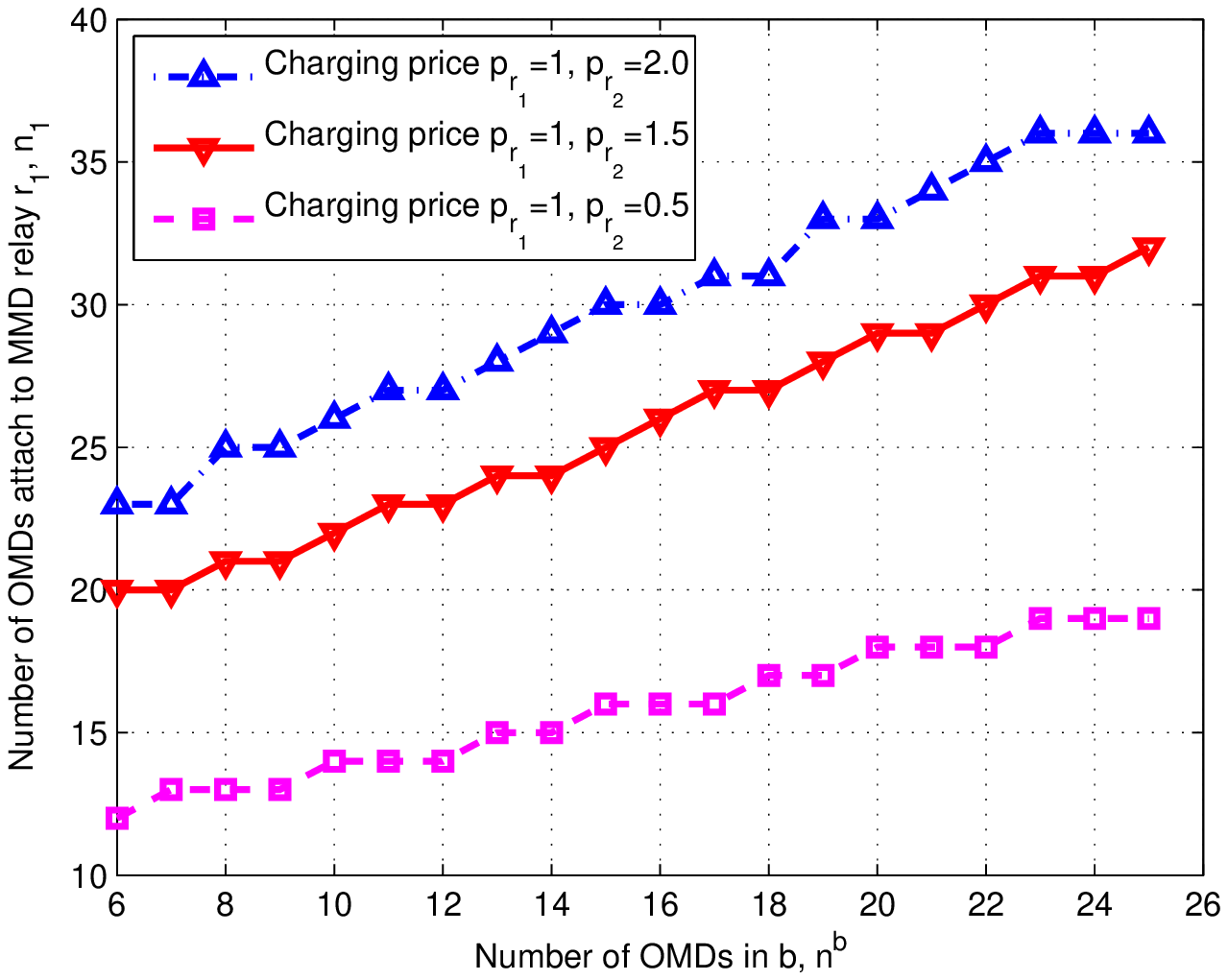}
	}
	\hspace{1mm}
	\subfigure[]{
		\label{fig:n1vsnbvb2}
		\includegraphics[width=0.2\textwidth]{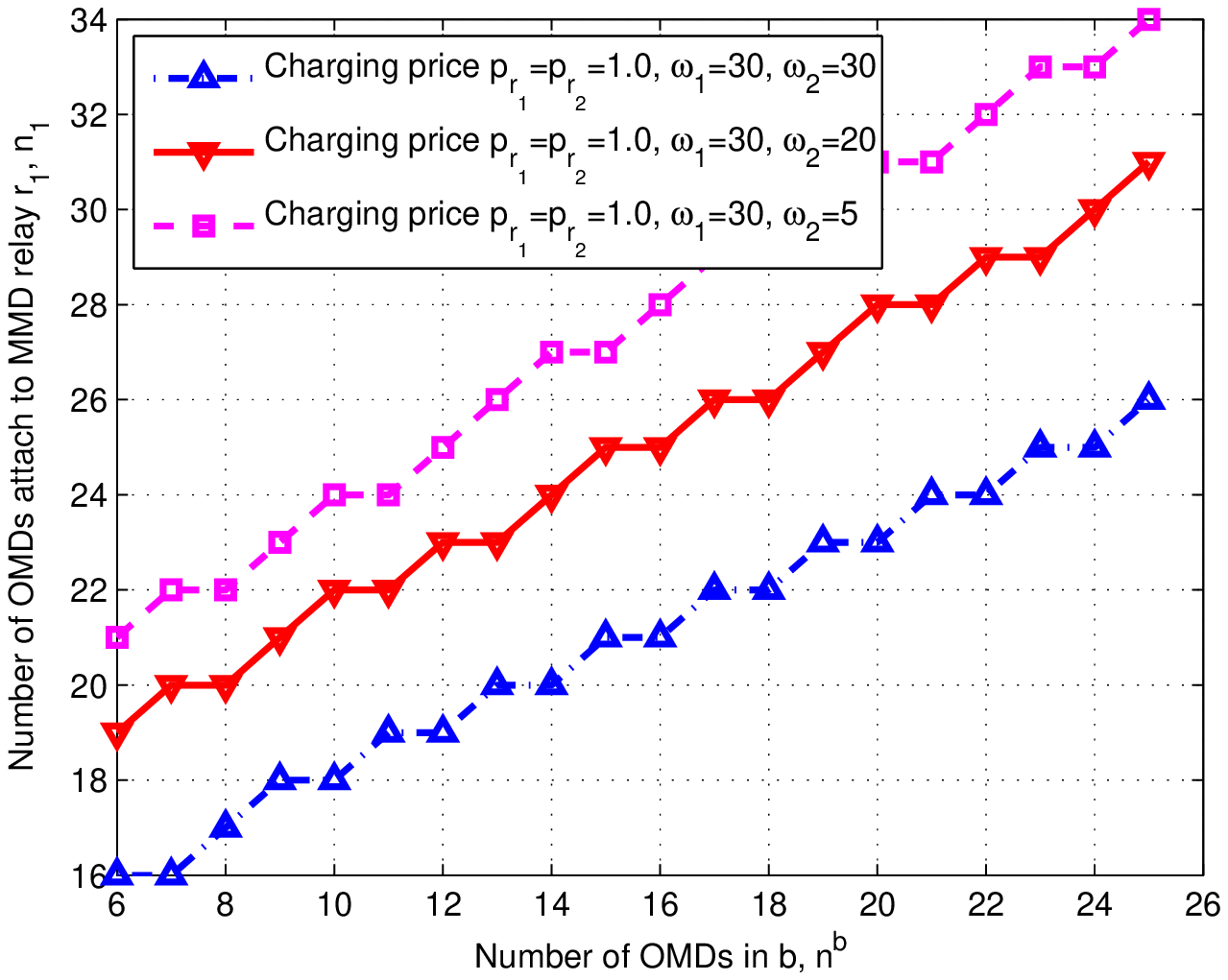}
	}
	\hspace{1mm}
	\subfigure[]{
		\label{fig:pr1pr2ne}
		\includegraphics[width=0.2\textwidth]{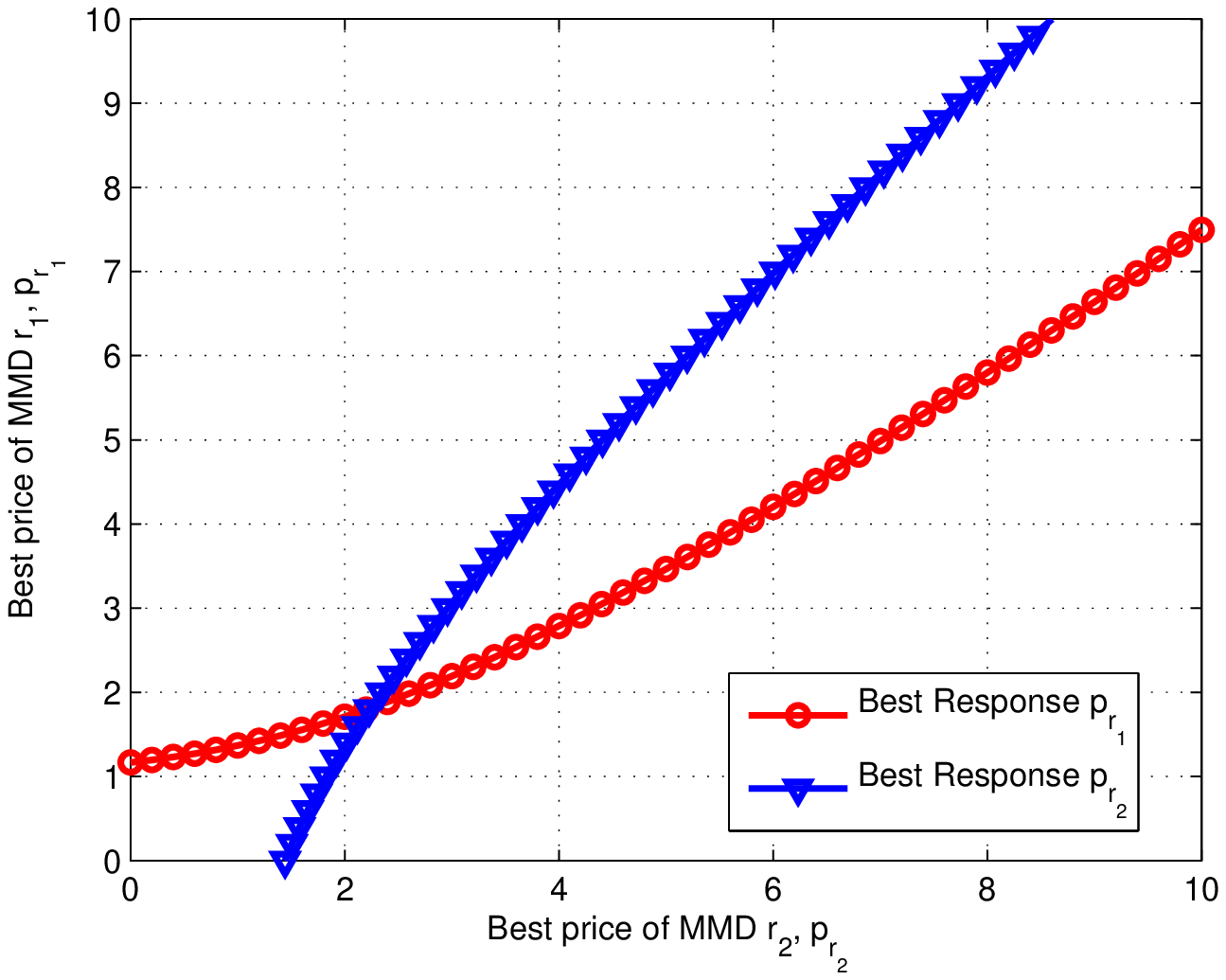}
	}
	\hspace{1mm}
		\subfigure[]{
		\label{fig:omega1omega2}
		\includegraphics[width=0.2\textwidth]{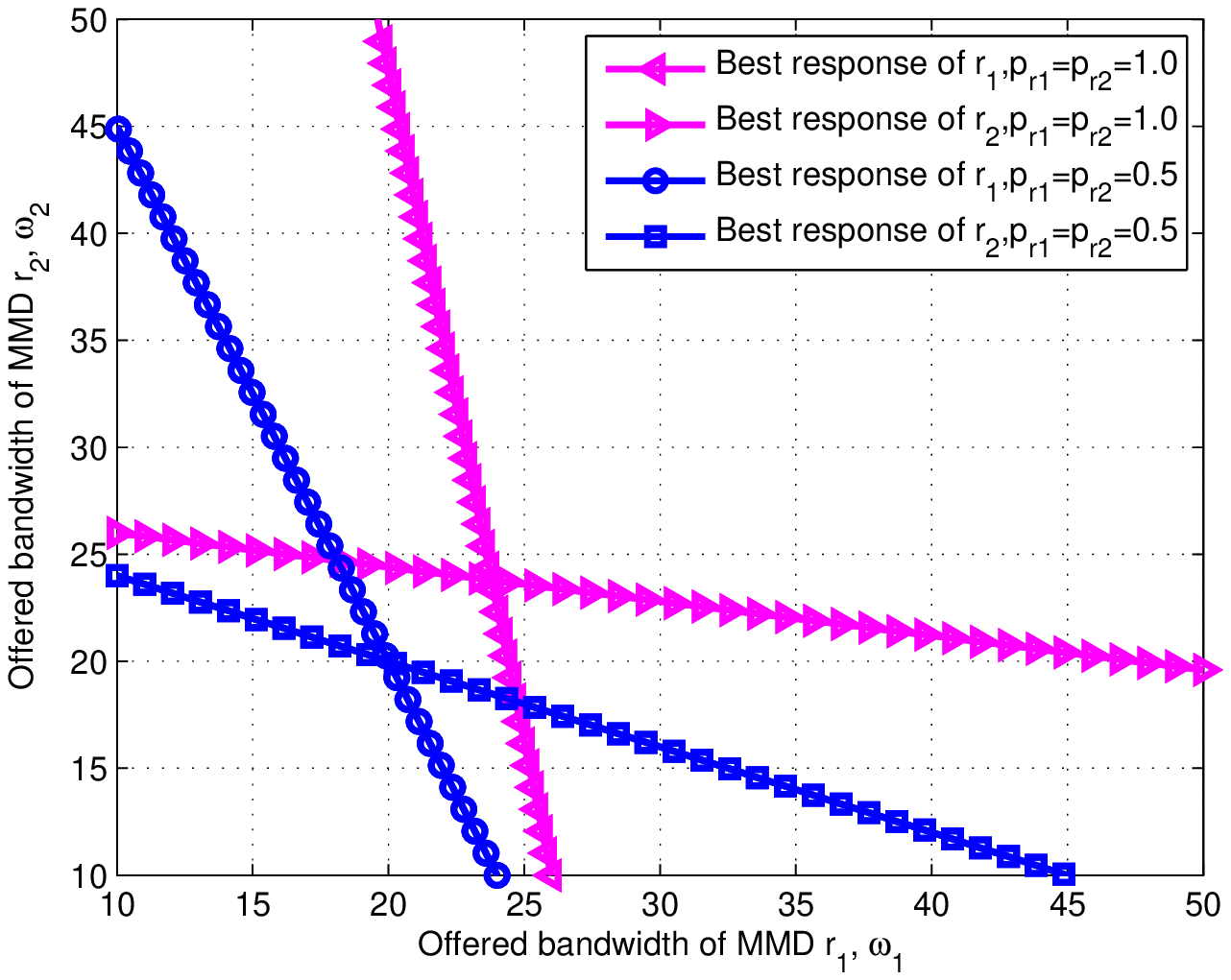}
	}
	\hspace{1mm}
	\small \caption{(a) Relationship between $n_1$ and $n^b$ with varying $p_{r_2}$ when $\omega_1=\omega_2=30$ and $n^a=25$ at equilibrium; (b) Relationship between $n_1$ and $n^b$ with varying $p_{r_2}$ when $\omega_1=30$ and $n^a=n^b=25$ at equilibrium; (c) Best responses of the two MMDs for the pricing strategy; (d) Best strategies of two MMDs with different charging prices.}
\end{figure*}

Further more, the basin of attraction consists of the replicator dynamics is the whole feasible region, that is $0<\pi_{1}^{a},\pi_1^{b}<1$. Fig. \ref{fig:Gutility2Nash} shows the convergence of the evolution of total utility in each group towards the equilibrium $(\pi_1^{a},\pi_1^{b})=(0.84,0.65)$ from the starting point $(0.73,0.50)$. In this experiment, we assume that both $r_1$ and $r_2$ offer the same bandwidth of $\omega_1=\omega_2=25$. Both $n^{a}$ and $n^{b}$ are equal to $30$. In the equilibrium, the utility of all OMDs in each group is close to the theoretical utility based on the observed value. In order to clearly present the results, we adjust the results of total utility divided by $100$, as shown in Fig. \ref{fig:Gutility2Nash}. It is observed that the total utility of group $a$ is higher than group $b$ when the equilibrium is achieved. That's because (i) the channel gains that result in the total utility for OMDs in group $a$ are higher than those in group $b$ and (ii) the ask price of $r_1$ is lower than that of $r_2$.
Next, we investigate the number of OMDs that attach to MMD $r_1$ with changing number of OMDs in group $b$. As shown in Fig. \ref{fig:n1vsnbvp2}, with increasing number of OMDs $n^b$, the number of OMDs choosing MMD $r_1$  grows accordingly. When the charging price of $r_2$, \ie $p_{r_2}$ grows from $0.5$ to $2.0$, $n_1$ increases due to the intention to choose a service provider with lower price. Further more, we study the relationship between $n_1$ and $n^b$ when MMDs $r_1$ and $r_2$ charge the same price at the evolutionary equilibrium. Let $p_{r_1}=p_{r_2}=1.0$, $\omega_1$ is fixed at $30$, as shown in Fig. \ref{fig:n1vsnbvb2}, when total number of OMDs in group $b$ is constant, $n_1$ decreases with the increase of $\omega_2$. That's because when the more bandwidth is offered by $r_2$, the less number of OMDs will choose $r_1$.

Fig. \ref{fig:pr1pr2ne} shows the best responses of the pricing strategy of the two MMD relays. In this experiment, MMD $r_1$ offers a bandwidth of $\omega_1=20$ and MMD $r_2$ offers a bandwidth of $\omega_2=40$. It is observed that both the best response strategy of $r_1$ and the best response strategy of $r_2$ grow in the Stackelberg game. Furthermore, for MMD $r_2$, the slope of the best response value is greater than $1$ whereas it is not the case for MMD $r_2$. That's because $r_2$ offers more bandwidth than $r_1$ does. The Nash equilibrium point is achieved when the best price strategy is $(p_{r_1}^{*},p_{r_2}^{*})=(1.8112,2.2341)$. To earn more profits, MMD $r_1$ should lease more bandwidth to OMDs.

To better demonstrate the MMD's best response with its counterpart's changing bandwidth and charging price, we compare the performance when the two-variable strategy profile $(\mathbf{\Omega}, \mathbf{P})$ is considered. In this example, we firstly fix the charging price of MMDs and derive the optimal bandwidth available to offer by MMDs. In this experiment, $\mu_{i.\omega}=1$, $\mu_{i,p}=0.5$ and $\Delta t=1$. As shown in Fig. \ref{fig:omega1omega2}, the charging prices are equal. When one MMD increases its offered bandwidth, more OMDs will choose it as their common resource provider while its counterpart will reduce the offered bandwidth to guarantee the utility of itself according to (\ref{eq:newU_i}). Accordingly, the OMDs will reach to the equilibrium following the similar phase plane of replication dynamics as shown in Fig. \ref{fig:phaseplane}. The Nash equilibrium can be achieved when both of the MMDs play their own best strategies with equal price. For example, when $p_{r_1}=p_{r_2}=1.0$, the equilibrium is at $(\omega_1^{*},\omega_2^{*})=(23.793,23.793)$. It can also be drawn from Fig. \ref{fig:omega1omega2} that when the charging prices increase, the best strategy for each MMD is to offer more bandwidth for OMDs so that it can gain more profits.

\begin{figure*}[htb]
	\centering
	\setlength{\abovecaptionskip}{-0.2cm}
	\setlength{\belowcaptionskip}{-1em}
	\subfigure[]{
		\label{fig:utilityna}
		\includegraphics[width=1.6in]{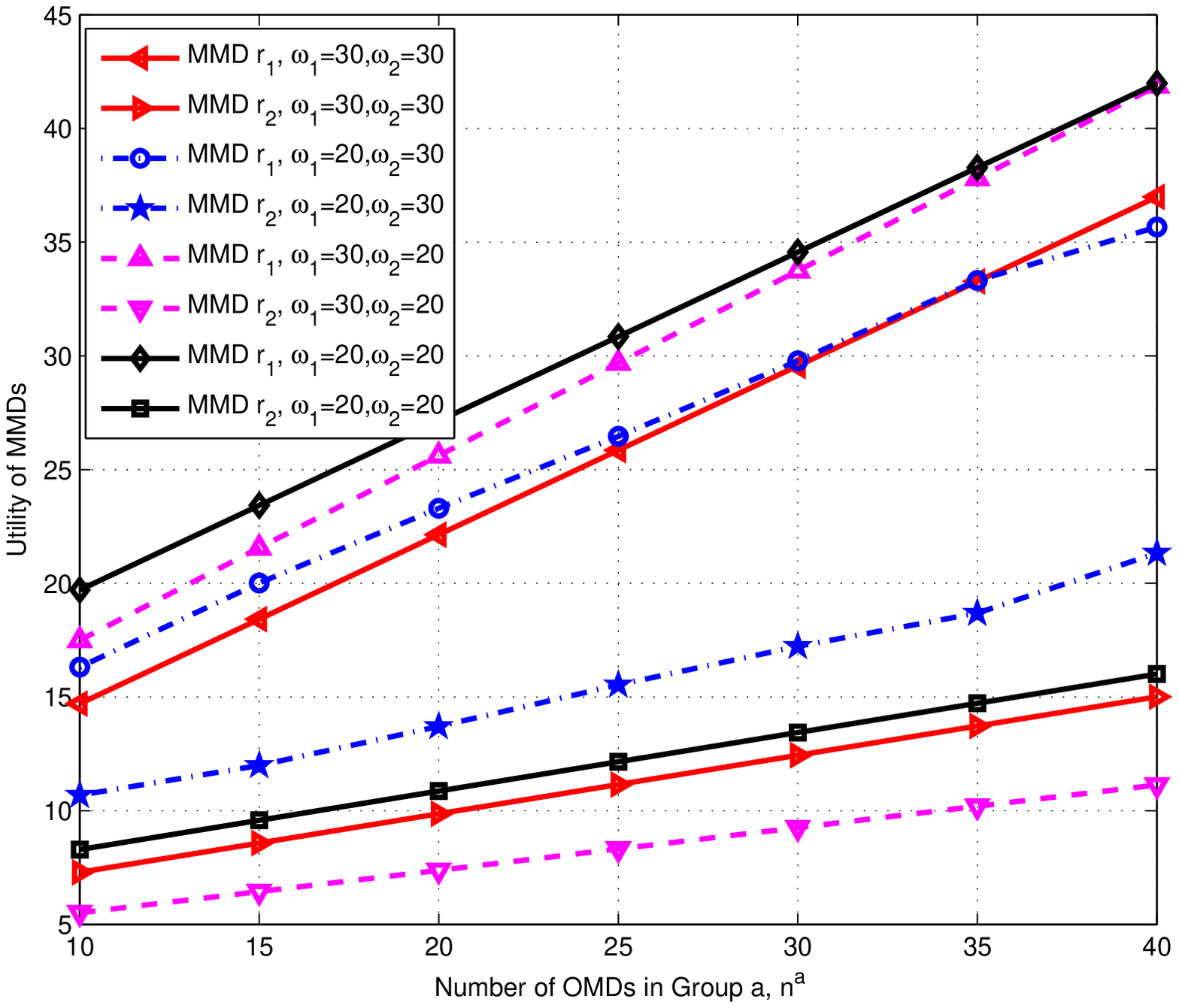}
	}
	\hspace{1mm}
	\subfigure[]{
		\label{fig:cc_dt}
		\includegraphics[width=1.6in]{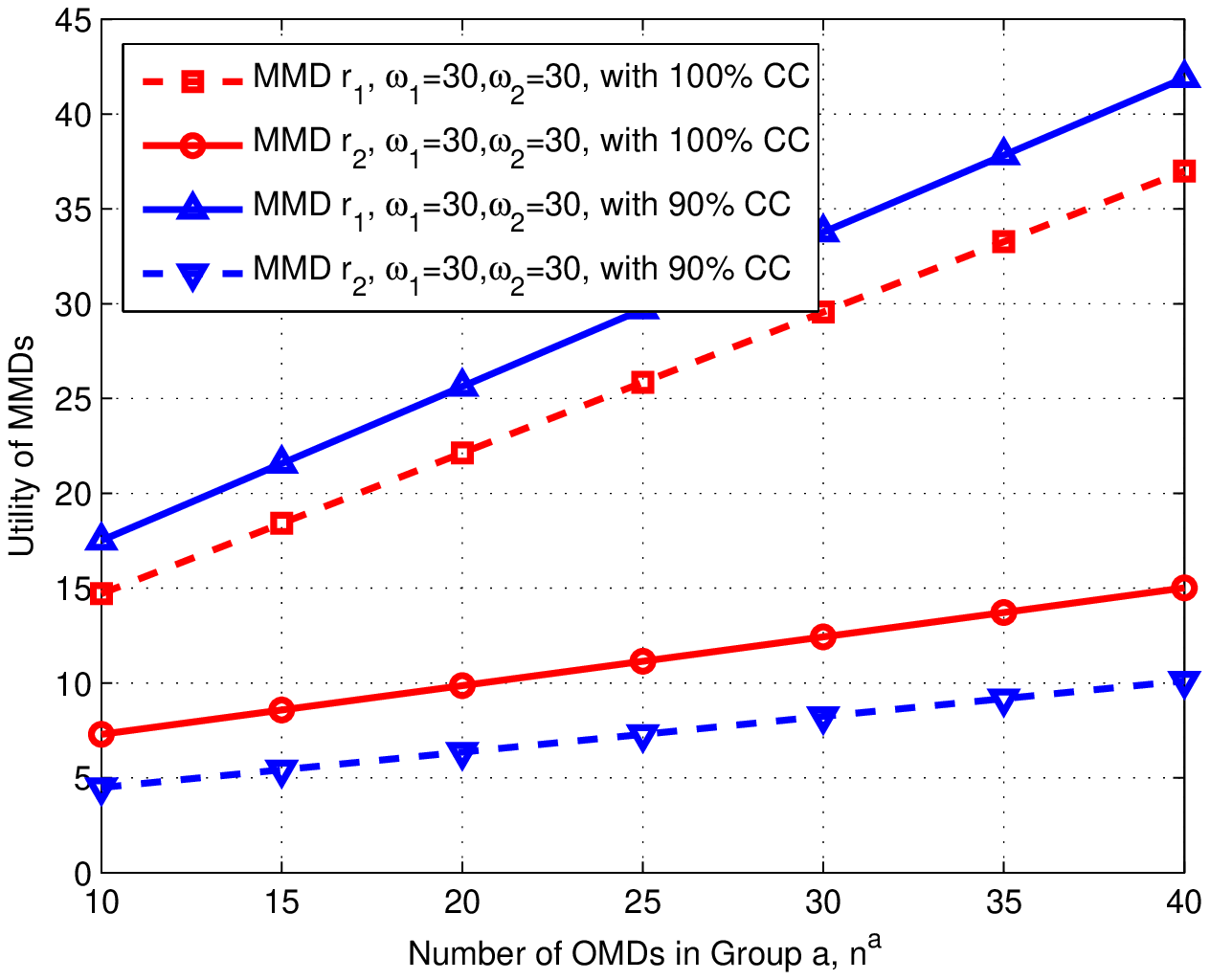}
	}
	\hspace{1mm}
	\subfigure[]{
		\label{fig:utilitycomapre}
		\includegraphics[width=1.6in]{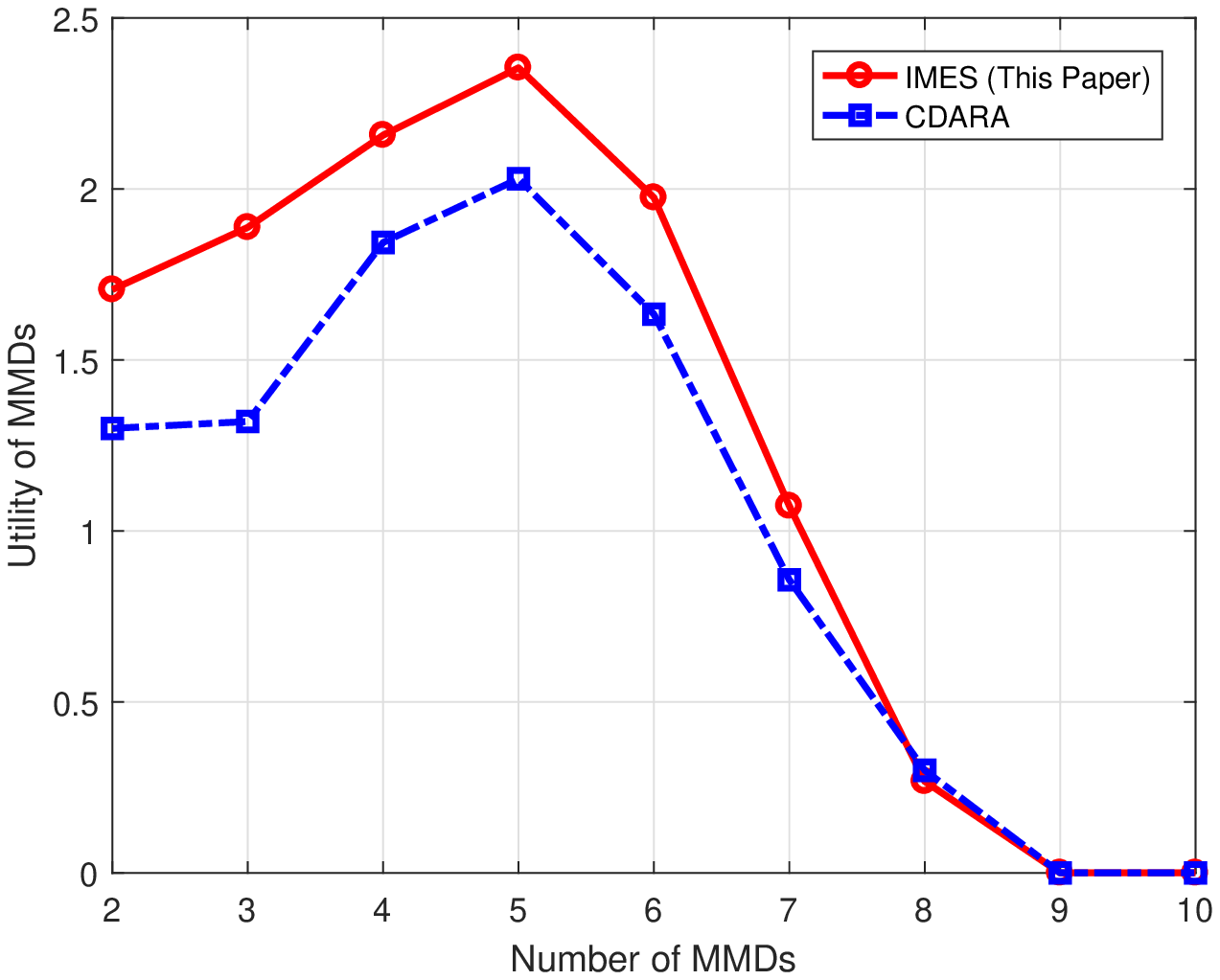}
	}
    \hspace{1mm}
    	\subfigure[]{
    	\label{fig:servdel_topo}
    	\includegraphics[width=1.4in]{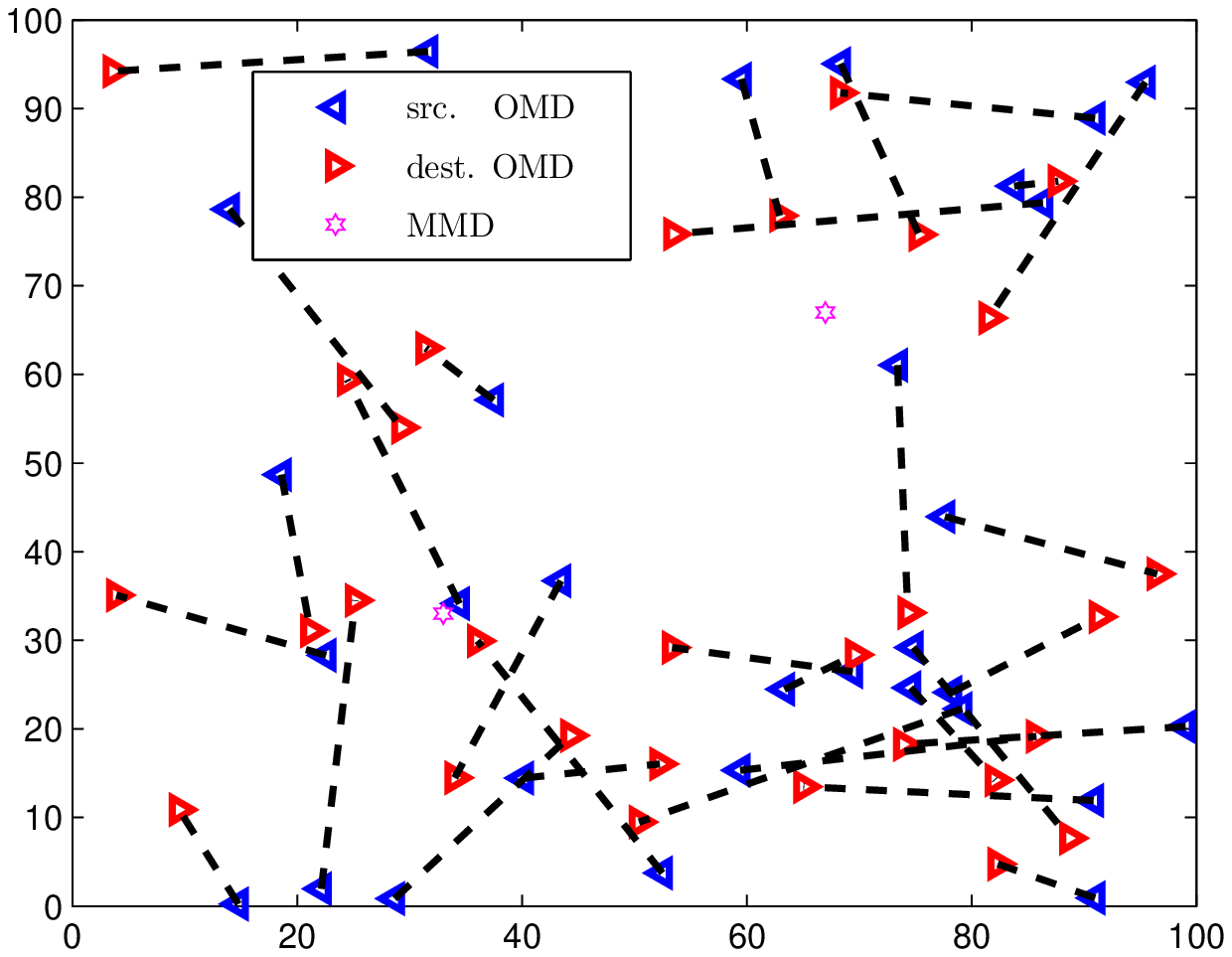}
    }
	\small \caption{(a) MMD's total utility with the number of OMDs in group $a$, $n^a$ at evolutionary equilibrium; (b) MMD's total utility with the number of OMDs in group $a$, $n^a$ at evolutionary equilibrium with changing proportion of cooperative communication pairs; (c) Utility of MMDs with the number of MMDs; (d) An example of network topologies in a plane of $100 \rm{m} \times 100 \rm{m}$.}
\end{figure*}

Fig. \ref{fig:utilityna} examines the utility of MMDs with the increasing number of OMDs in group $a$. When the bandwidth $\omega_2$ is fixed at $\omega_2=20$, the utilities of both MMDs increase when $\omega_1$ drops from $30$ to $20$. That's because when $\omega_1$ becomes smaller, a lower price will be at the equilibrium, which can attract more OMDs (as previously explained of Fig. \ref{fig:omega1omega2}). We have the similar findings when $\omega_2$ is fixed at $\omega_2=30$ as shown in Fig. \ref{fig:utilityna}.

Next, we analyze the relationship between net utility of MMDs and the number of OMDs in group $a$. Let $c_1=0.5$, $c_2=0.1$, $n^b=30$. Generally, the utility of MMDs grows with the number of OMDs in group $a$. For MMD $r_1$, when $\omega_1=30$ and $n^a$ is fixed, with $\omega_2$ decreasing from $30$ to $20$, the utility of MMD $r_1$  increases because there are more OMDs choose to switch from $r_2$ to $r_1$ due to more resource offerings by $r_1$. In Fig. \ref{fig:cc_dt}, when cooperative communication is adopted, the MMD acts not only as the bandwidth seller, but also acts as the intermediate relay, which differentiates the final utility function of MMDs. Take Fig. \ref{fig:cc_dt} as an example again. When there are only $90\%$ OMDs conducting cooperative communication (CC) with MMD $r_1$ in group $a$, the utility difference between $r_1$ and $r_2$ is larger than that of all nodes using cooperative communications, that can be explained by the changing values of $k_{\omega} \tau_{ij}$ in (\ref{eq:new_CR}).

To study the influence of more than two MMDs with constant number of OMDs, we modify the algorithm CDARA\cite{samimi2016combinatorial} so that it can fit for the case of only one kind of resource demands of buyers. We compare the proposed IMES with the modified CDARA scheme. Each newly added MMD has the initial bandwidth of $40$ and $p_{r_j}=0.2$, where $j>2$. The valuation of each OMD with CDARA scheme is calculated by their original position and the two groups can be divided into multiple sub-groups with each MMD acts as the group leader. As shown in Fig. \ref{fig:utilitycomapre}, with the growing number of MMDs, the total utility of MMDs grows first and decreases fast when the number of MMDs is larger than $5$ and finally reaches to zero utility. That's because the more MMD relays, the more user demands can be satisfied. However, when the number of MMD relays reaches to the limit of $5$, the limited number of OMD users in one single sub-group cannot afford the MMD relay, thus the total utility will decrease and finally reaches to zero utility. On average, IMES outperforms the CDARA scheme by about $20.37\%$.

\subsection{Service Delay}
\label{subsec:service_delay}

To study service delay, we compare our proposed incentive based method IMES with a randomized method (named as Rand). Regarding to Rand, an OMD randomly selects one of the MMDs for CD2D communication if CD2D communication is better than direct D2D communication. With regard to our proposed IMES, details can be found in Section \ref{sec:imes}.

We first consider an example network in a square area of $100 \rm{m} \times 100 \rm{m}$, as shown in Fig. \ref{fig:servdel_topo}, where the left-pointing triangle represents source OMD and the right-pointing triangle stands for destination OMD.  During the simulation, MMDs always have extra bandwidth that can be shared with OMDs. If an OMD cannot finish the transmission in the current round of TDMA slots, it will be served within the following rounds. Therefore, we define the \emph{service delay as the average transmission time when an OMD finishes sending messages to its destination}.

\begin{figure*}[t]
	\centering
		\setlength{\belowcaptionskip}{-1em}
	\subfigure[Average service delay versus the number of OMDs in a $100 \rm{m} \times 100 \rm{m}$ network ]{
		\label{fig:sd_omd_omegav}
		\includegraphics[width=1.6in]{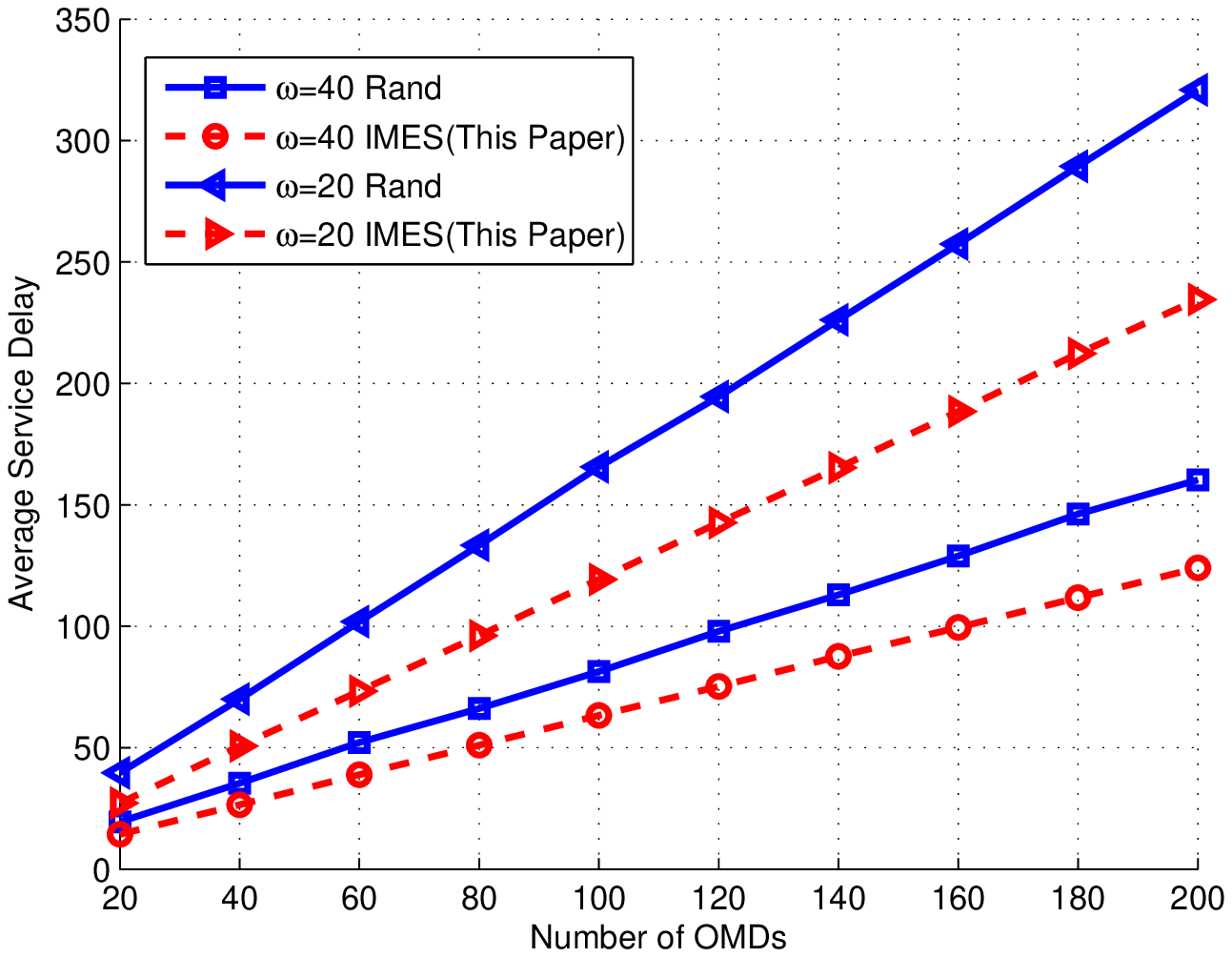}}
	\hspace{1mm}
		\subfigure[Average service delay versus the number of MMDs in a $100 \rm{m} \times 100 \rm{m}$ network]{
		\label{fig:servdelaymmd} 
		\includegraphics[width=1.6in]{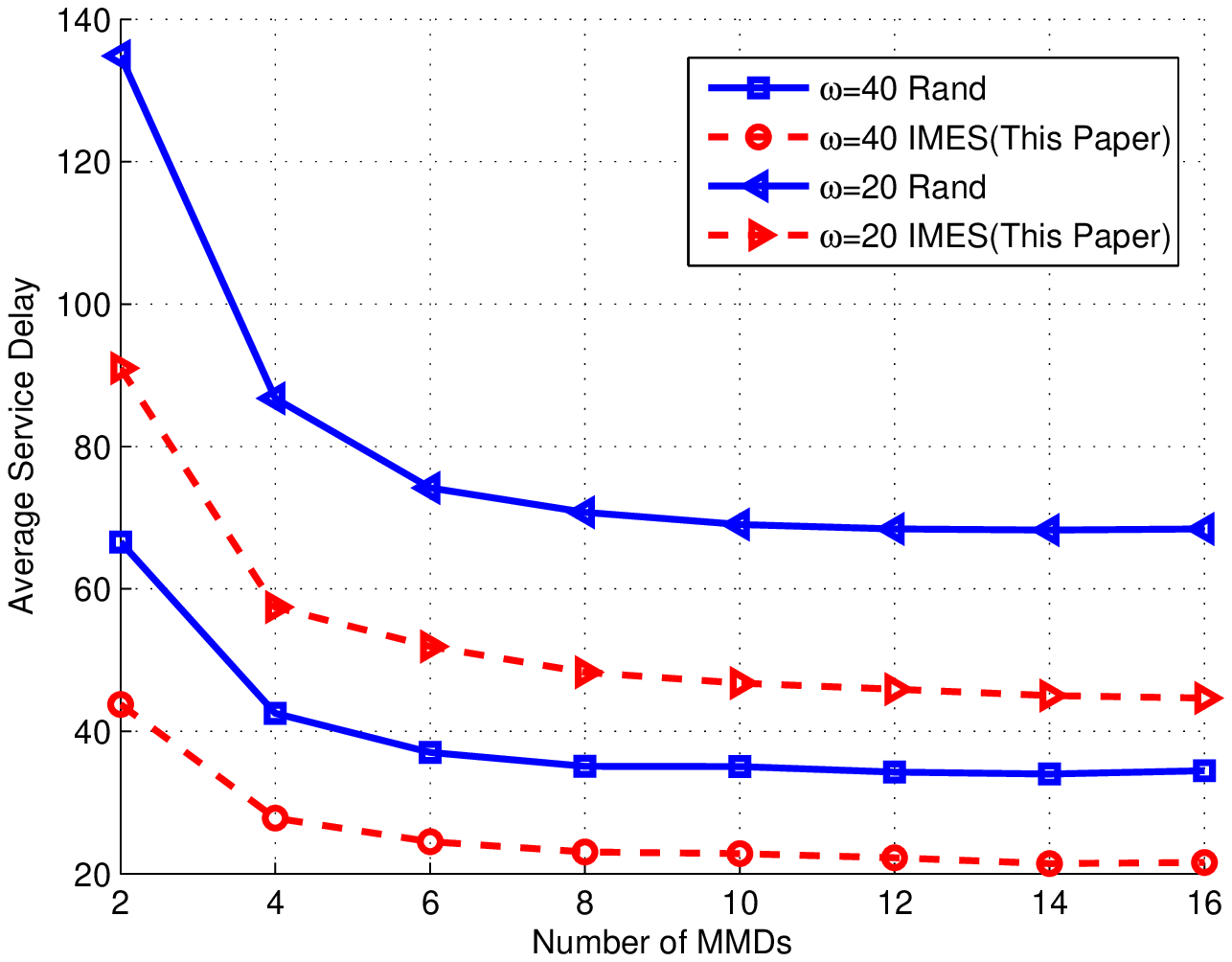}}
	\hspace{1mm}
	\subfigure[Average service delay versus various network sizes with the fixed number of MMDs and OMDs]{
		\label{fig:avsdas} 
		\includegraphics[width=1.6in]{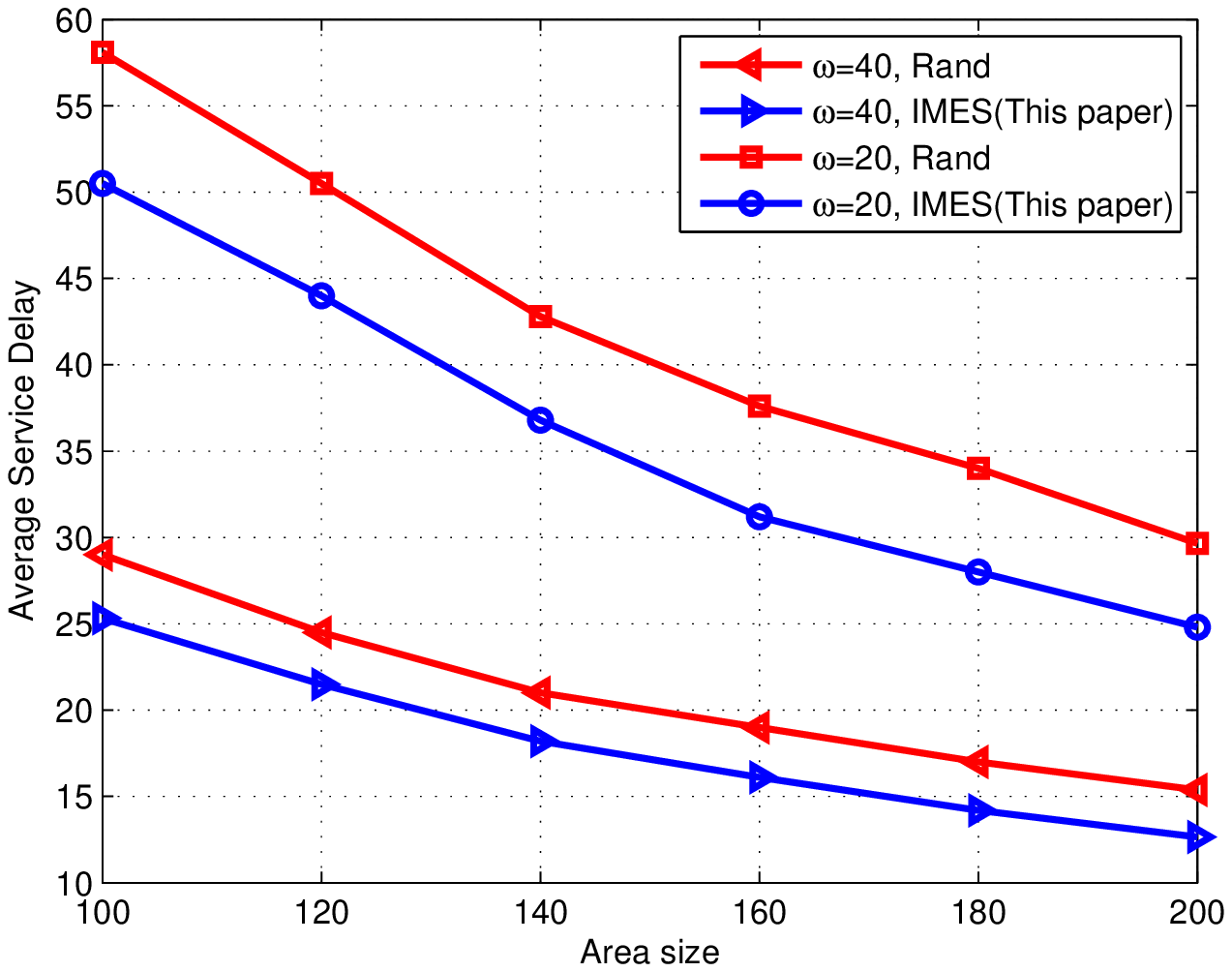}}
	\hspace{1mm}
	\subfigure[Average service delay versus various network sizes with changing number of MMDs]{
		\label{fig:cntomgvasnmmd} 
		\includegraphics[width=1.6in]{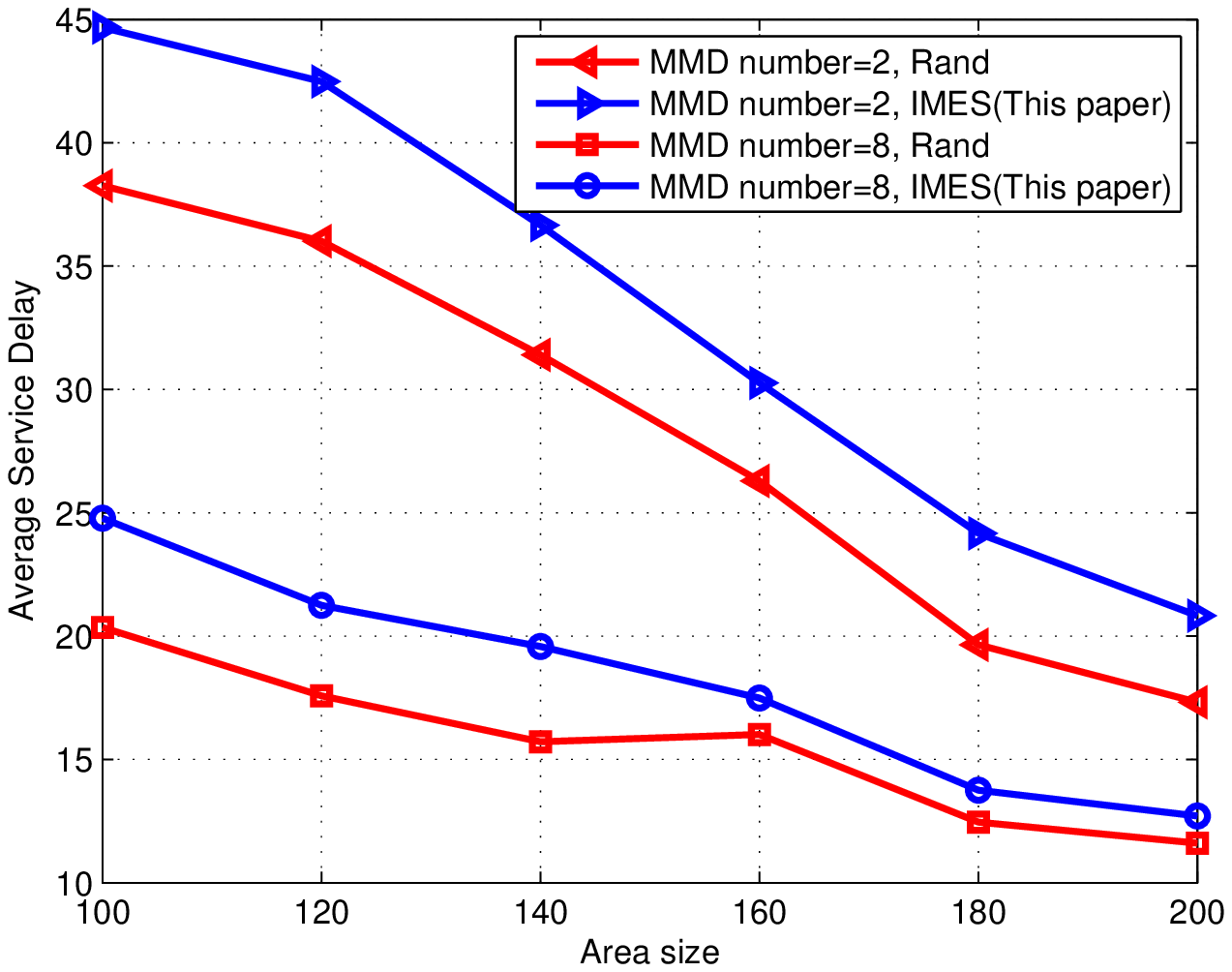}}
	\caption{Performance of average delay}
	\label{fig:subfig} 
\end{figure*}

We exam the impact of both the number of OMDs and the number of MMDs on service dealy. Fig. \ref{fig:sd_omd_omegav} shows the average service delay versus the number of OMDs when the number of MMDs is fixed at $|\mathcal{R}|=3$. It is shown in Fig. \ref{fig:sd_omd_omegav} that the average service delay increases with increased number of OMDs in both IMES and Rand schemes. That's because more OMDs cause higher traffic loads, which results in the higher service delay. Besides, Fig. \ref{fig:sd_omd_omegav} also shows that our proposed IMES always has lower service delay than Rand scheme, implying the effectiveness of our IMES. On average, IMES is more than 25\% times better than the randomized method. Moreover, when the number of OMDs is fixed, the average service delay drops quickly with the increased bandwidth. For example, when there are 120 OMDs, the available bandwidth is $\omega=20$, the average delay is about $200$ for Rand scheme and $150$ for IMES while the delay is only $100$ for the random method and $75$ for IMES. This observation implies that increasing the bandwidth can significantly reduce the service delay.

Fig. \ref{fig:servdelaymmd} shows the relationship between the number of MMDs and the average service delay. We fix the number of OMDs at $50$ in the network. It is shown in Fig. \ref{fig:servdelaymmd} that the average service delay decreases fast with the increased number of MMDs when the number of MMDs is below $8$, implying that increasing the number of MMDs can significantly reduce the service delay. However, the average service delay reaches a constant value when the number of MMDs is greater than 8. That is because the number of OMDs served by each MMD reaches a saturated value with the fixed bandwidth, implying that adding more MMDs will not bring any benefits in this case. Besides, it is also shown in Fig. \ref{fig:servdelaymmd} that our proposed IMES always has lower average service delay than the Rand. On average, IMES is about $32.8\%$ better than Rand when the available bandwidth for each MMD is $\omega=40$. When $\omega=20$, the delay reduction gain is about $36.7\%$.

The results in both Fig. \ref{fig:sd_omd_omegav} and Fig. \ref{fig:servdelaymmd} are based on a network with the fixed area $100 \rm{m} \times 100\rm{m}$. We next investigate the impacts of network area on the service delay. In the second set of simulations, we increase the network area from $100 \rm{m} \times 100 \rm{m}$ to $200 \rm{m} \times 200 \rm{m}$. Note that we fix the number of OMDs at $40$ and the number of MMDs at $8$. It is shown in Fig. \ref{fig:avsdas} that the average service delay decreases with the increasing of network area. That's because the traffic congestion is alleviated when network becomes sparse. Besides, our proposed IMES can reduce average service delay by at least $20\%$ compared with Rand scheme with consideration of network area size influence.

We then conduct another set of simulations to investigate impacts of the number of MMDs with consideration of the changing network area size. In particular, we deploy $60$ OMDs in the network. It is shown in Fig. \ref{fig:cntomgvasnmmd} that increasing the number of MMDs can greatly decrease the average service delay, implying that a better QoS is achieved with the growth of MMDs.
\section{Conclusion and Future Remarks}\label{sec:conclude}

In this paper, we have proposed a novel multi-homing cooperative D2D network, by combining the ordinary mobile devices and multi-homing mobile devices. We have contributed a two-level evolutionary game for such MH-CD2D networks, to deal with one of the challenges of the network, \ie, the joint bandwidth relay allocation problem. We have provided analysis for the procedure to select the best MMDs utilizing the replica dynamics. We have also proposed algorithms to maximize the utilities of the OMDs by introducing price competitions into the scheme, instead of keeping the price unchanged in the traditional auction mechanisms. Moreover, we have investigated the existences of the evolutionary equilibrium in the proposed MH-CD2D networks. Extensive simulation results demonstrate that the equilibrium can be achieved and the best response price of one MMD increases with the best price of the other MMD in the Stackelberg game. The utility of MMDs increases with the number of OMDs in each OMD group at the evolutionary equilibrium. For service delay, we take the influence of network area into consideration. Following the most existing works, we compare our algorithms with the randomized scheme. Experimental results show that the proposed scheme IMES is able to improve the randomized scheme by more than $25\%$ in terms of service delay. With IMES, the utilities of MMDs and OMDs can be enhanced. Meanwhile, the fairness of OMD within the same group and between different groups can be ensured. When the prices  of both MMDs increase towards the equilibrium (shown in Fig. \ref{fig:pr1pr2ne}), the two MMDs tend to offer more bandwidth for OMDs (shown in Fig. \ref{fig:omega1omega2}). Therefore, IMES has shown its advantage to motivate bandwidth sharing as well as to ensure fairness.

Our future work includes addressing the problem when OMD users move between different macro-cells or small cells. We also plan to investigate the case when one MMD node can serve multiple OMDs and one OMD can compete for multiple MMDs. In addition, our future work also covers interference control between neighboring MMDs and the study for general cases with different mobility models.



\end{document}